\documentclass[letterpaper, 10 pt, conference]{ieeeconf}

\IEEEoverridecommandlockouts                              
\usepackage[utf8]{inputenc}
\usepackage[T1]{fontenc}

\usepackage{graphics} 
\usepackage{epsfig} 
\usepackage{mathptmx} 
\usepackage{mathptmx} 

\usepackage{amsmath} 
\usepackage{amssymb,amsthm}  
\usepackage{color}

\usepackage{tikz}
\usetikzlibrary{calc,arrows,positioning,shapes}

\newtheorem{thm}{Theorem}
\newtheorem{lem}{Lemma}
\newtheorem{ass}{Assumption}

\newtheorem{rem}{Remark}

\usepackage{tikz,pgfplots}
\pgfplotsset{compat=newest}
\usepackage[list=true, font=large, labelfont=bf, 
labelformat=brace, position=top]{subcaption}
\usepgfplotslibrary{external}

\usetikzlibrary{calc,arrows,positioning,shapes}
\usetikzlibrary{external}

\definecolor{mycolorred}{rgb}{0.85000,0.32500,0.09800}%
\definecolor{mycolorblue}{rgb}{0.00000,0.44700,0.74100}%

\title{\LARGE \bf
Control of Unknown (Linear) Systems with Receding Horizon Learning 
}

\author{C. Ebenbauer$^{1}$ and F. Pfitz$^{1}$
and S. Yu$^2$
\thanks{$^{1}$University of Stuttgart,
Germany {\tt\small ce@ist.uni-stuttgart.de,
fabian.pfitz@ist.uni-stuttgart.de},
$^2$Jilin University,
China {\tt\small shuyou@jlu.edu.cn}.
The work is supported partially by the National Nature Science Foundation of China (No.61520106008).
        }}

\begin{document}

\onecolumn

\maketitle

\begin{abstract}
A receding horizon learning scheme is proposed to transfer the 
state of a discrete-time dynamical control system to zero without the need of a system model. Global state convergence to zero is proved for the class of stabilizable and detectable linear time-invariant systems, 
assuming that only input and output data is available and an upper bound of the state dimension is known. 
The proposed scheme consists of a receding horizon control scheme  and a proximity-based estimation scheme to estimate and control the closed-loop trajectory.
Simulations are presented for linear and
nonlinear systems.
\end{abstract}

\section{Introduction}

Currently, a lot of research effort is centered around the interplay between control, learning, and optimization. This is driven by extensive research initiatives in artificial intelligence, by the steadily increasing online computing power, and by the wish to build autonomous and intelligent systems in all sorts of application domains.
From these developments, a renewed interest in the control of systems 
where no system model is known,
or where the model involves large uncertainties
has
emerged under the banner of learning-based or data-based control.
Traditionally, this is a subject of adaptive control.
In this vein, we address a classical problem from adaptive control,
namely the stabilization of completely unknown linear time-invariant
discrete-time control systems.
We aim for a solution that utilizes online optimization and (past and future) receding horizon data, and that provides convergence guarantees.
To this end, we propose a scheme that involves estimation, prediction, and feedback control for unknown systems, which we have subsumed  under the term learning in the title of this work.

The  literature on the (adaptive) stabilization of unknown systems is huge.
Many different solution approaches exist
in the adaptive control literature ranging from model-free approaches  to model-based approaches \cite{Ben-16, Tao-14, Goo-09}.
The control of unknown systems has also been studied in the area of optimal control and receding horizon control for quite some time,
see e.g. \cite{Fel-60,Mos-95,Pet-82,Ber-96}. 
Work that is related to our work in the sense that similar problems and challenges are addressed are for example \cite{Mos-95} (Chapter 3) and \cite{Bem-94},
in which an input-output stabilizing receding horizon control was proposed and combined with a multi-step prediction method for parameter model identification. 
Further, in the works \cite{Bit-Bol-Cam-90} convergence of a recursive least-squares identification algorithm was addressed with incomplete excitation and convergence of an adaptive control scheme was proven
under the assumption that the system has an asymptotically stable zero dynamics.
Also related is for example the publications \cite{Ter-19},
in which a framework based on a robust model predictive scheme and
identification of a multi-step prediction model was proposed and convergence and feasibility for stable systems were shown.
A common approach when controlling completely unknown systems is based on a combination of a control scheme (such as pole-placement) and an
online estimation or identification scheme (such as recursive least-squares). 
Hereby, models are estimated and updated in real-time based on the measured input-output data and these
models are utilized in the control scheme.
A major challenge when using this so-called certainty equivalence approach to stabilize unknown (and unstructured) systems is the \emph{loss of stabilizability problem}, i.e. how to ensure in a computationally  efficient way that the estimated models are, for example, stabilizable so that adaptive pole-placement can be applied. 
See for example \cite{Bit-06,Hes-Lib-Mor-03,Mor-92,Man-Tu-Rec-19,Pra-Cam-98} and
references  therein on this topic.
More recent related research on (partially) unknown systems and receding horizon control are for example discussed in
\cite{Ngu-20,Ade-09,Tap-Fra-20,Luc-18,Korda17,Mue-16,Lim-17,Bec-18,Ber-20,Hew-20,Sch-Gha-Tri-Ebe-18,Pap-Ros-Bor-20,May-14}, to mention only a few out of the rapidly growing literature.
A complete review of the state of the art of controlling (partially) unknown systems with receding horizon schemes is out of the scope of this work.
The proposed approach in this paper  is based on the classical certainty equivalence implementation. However, in contrast to the existing literature, we 
provide a fully online optimization-based solution with
provable convergence for completely unknown linear systems.
For example, we do not assume that the linear system is persistently 
excited, which is often assumed in order to identify (directly or indirectly) 
a system model. Moreover, we do not assume that the state can be measured
nor that the system is stable or controllable.
Under these minimal assumptions \cite{Mar-85,Mar-93,Hel-91}, the satisfaction of state or input constraints is not feasible and we do therefore not consider  constraints 
as it is usually done in the receding horizon (predictive control) literature. 
Nevertheless, the control of unknown linear systems
is an important benchmark problem, and, to the best of our knowledge, a receding horizon approach that provably ensures state convergence under these assumptions has not been reported in the literature.
In particular, the contributions of this work are as follows.
We propose an online optimization scheme which builds on a
receding horizon control scheme 
and an estimation scheme. 
The receding horizon control scheme 
is based on a novel model-independent terminal state weighting
in the sense that the objective function and the terminal cost can be chosen independently of a (not necessarily controllable) system model. 

The estimation scheme is based on a modified proximal minimization algorithm that guarantees convergence of the estimated quantities,  and does not require that the closed loop system is persistently excited. 
A  characteristic feature of the approach is that the estimated quantities 
do not correspond to a (or to the "true") system model but rather to a signal model (time series or signal predictor) of the closed loop trajectory. 
The overall computational online effort of the proposed scheme is rather low and requires essentially the solution of (least-squares) regression problems.
Finally, the proposed scheme is also
applicable to nonlinear systems as demonstrated by simulations.

\section{Problem Statement}
\label{problemstatement}

Consider the discrete-time linear time-invariant system
\begin{align}
\label{sys}
\begin{split}
z(k+1)&=Fz(k)+Gv(k)\\
y(k)&=Hz(k)
\end{split}
\end{align} 
with state $z(k) \in \mathbb{R}^n$, input $v(k) \in \mathbb{R}^q$ and output $y(k) \in \mathbb{R}^p$ at time instant $k \in \mathbb{N}$.

\begin{ass}
\label{ass1ground}
We assume that $(F,G)$ is stabilizable
and $(F,H)$ detectable.
Furthermore, we assume $F,G,H$ are unknown,
that an upper bound $m \ge n$ of the state dimension is known,
and that only past input and output data is available.
\end{ass}

Under Assumption \ref{ass1ground},
we aim to define a scheme
which guarantees 
for any initial state $z(0)$ 
that the system state $z(k)$ of \eqref{sys} converges to zero as time index $k$ goes to infinity.
We develop the scheme in three steps. 
In a first step, in Section \ref{rhcpart},  we develop a stabilizing,
model-independent receding horizon control scheme
based on asymptotically accurate predictor maps for the closed loop trajectory.
In a second step, in Section \ref{estimationpart}, we develop a 
proximity-based estimation scheme to obtain the asymptotically accurate predictor maps in terms of a so-called signal model.
Section \ref{rhcpart} and  \ref{estimationpart} 
are independent of each other and also the contributions therein. In a third step, in Section \ref{overallpart}, the control scheme and
the estimation scheme are combined in a proper way to solve the stated problem.
In Section \ref{secsim}, we provide simulation results.
All proofs can be found in the Appendix.

\section{A Model-independent Receding Horizon Control Scheme}
\label{rhcpart}

\subsection{Problem Setup}

Consider the system
\begin{align}
\label{sysmpc}
\begin{split}
x(k+1)&=Ax(k)+Bu(k)\\
\end{split}
\end{align} 
with $x(k) \in \mathbb{R}^n$, $u(k) \in \mathbb{R}^q$.
Further, consider the following optimization problem
\begin{align}
\label{mpc1}
\begin{split}
     V_1(x,p_1)=\min  ~&~  \sum\limits_{i = 0}^{N - 1}  x_i^{\top} Q x_i + u_i^{\top} R u_i  +  \frac{\Gamma(x)}{\epsilon} x_N^{\top} Q_N x_N\\
    \text{s.t. } ~&~ x_{i} = P_i(k,x_0,u_0,...,u_{i-1}), ~i=1...N \\
                 ~&~ x_0 = x 
\end{split}    
\end{align}
with  $p_1^\top=[k,\epsilon]$, $N,k \in \mathbb{N}$, $\epsilon>0$,
$\Gamma: \mathbb{R}^{n} \rightarrow \mathbb{R}$ nonnegative.
The decision variables are $u_i \in \mathbb{R}^{q}$, $i=0...N-1$ and $x_i\in \mathbb{R}^{n}$, $i=0...N$
and we refer to $x$ and $p_1$ as parameters.
The map $P_i:\mathbb{N} \times \mathbb{R}^{n} \times \mathbb{R}^{iq} \rightarrow \mathbb{R}^{n}$
is an $i$-th step-ahead state (or signal) predictor.
We denote the value of the objective function for some $u_{i}$, $i=1...N-1$ with $V_0(x,p_0,u_0,...,u_{N-1})$
(since the variables $x_i$  are determined by $u_i$'s and $x$)
and an optimal solution  
is denoted by $u_i(x,p_1)$, $i=0...N-1$,
$x_i(x,p_1)$, $i=1...N$.
If in \eqref{mpc1} we choose $x$ to be the state of \eqref{sysmpc} at time instant $k$, i.e.
$x=x(k)$ and if we choose $p_1=p_1(k)$ at time instant $k$ for a given sequence $\{p_1(k)\}_{k \in \mathbb{N}}$, then we refer to a mapping
\begin{align}
x(k) \mapsto u_0(x(k),p_1(k))
\end{align}
as the \emph{receding horizon control policy} defined by \eqref{mpc1}
and we call \eqref{mpc1} and \eqref{sysmpc} \emph{closed loop}, if
 in \eqref{sysmpc} $u(k)=u_0(x(k),p_1(k))$.
We impose the following assumptions.
\begin{ass}
\label{ass1}
$(A,B)$ in \eqref{sysmpc} is stabilizable
and the state can be measured.
\end{ass}

\begin{ass}
\label{ass1a}
The prediction horizon satisfies $N \geq n=\mathrm{dim}(x(k))$ and  $Q>0,R>0,Q_N>0$, $\Gamma:\mathbb{R}^n \rightarrow \mathbb{R}$ positive definite.
\end{ass}

\begin{ass}
\label{ass0}
(a) For any $k \in \mathbb{N}$, we assume that the state predictor maps $P_i$, $i=1...N$
have the following linear structure
\begin{align}
\label{predictor}
\begin{split}
  P_{i}(k,x_0,u_0,...,u_i) = A_{i}(k)x_0+\sum_{l=0}^{i-1} B_{i-1-l}(k) u_{l}
\end{split}  
\end{align}
where $\{A_i(k)\}_{k \in \mathbb{N}}$, $i=0...N$ with $A_0(k)=I$, and $\{B_i(k)\}_{k \in \mathbb{N}}$, $i=0...N-1$, 
$A_i(k) \in \mathbb{R}^{n \times n}$, $B_i(k) \in \mathbb{R}^{n \times q}$, are convergent matrix sequences, i.e.
\begin{align}
\label{p2}
\begin{split}
\lim_{k \rightarrow \infty} A_i(k)=\hat A_i, ~\lim_{k \rightarrow \infty} B_i(k)=\hat  B_i.
\end{split}
\end{align}
(b) Moreover,  for any $x(0) \in \mathbb{R}^n$
the state predictor maps $P_i$ along the trajectory $x(k)$, $k \in \mathbb{N}$, of the closed loop
\eqref{mpc1} and \eqref{sysmpc} with $s_l:=u_l(x(k),p_1(k))$ 
(or of the closed loop \eqref{mpc3} and \eqref{sysmpc}
with $s_l:=\nu_l(x(k),p_3(k))$)
satisfy for $0 \le i+j\le N+1$
\begin{align}
\begin{split}
P_j(k,P_i(k,x(k),s_0,...,s_{i-1}),s_{i},..,s_{i+j-1}) = \\ P_i(k,P_j(k,x(k),s_0,...,s_{j-1}),s_{j},..,s_{i+j-1})=\\P_{i+j}(k,x(k),s_0,...,s_{i+j-1}).
\end{split}\label{p4}
\end{align}
Further, we assume that state predictor maps
predict asymptotically accurate with respect to system \eqref{sysmpc}
in the sense that we have for any $k \in \mathbb{N}$, $i=0...N$
\begin{align}
A_{i}(k) x(k)+\sum_{l=0}^{i-1} B_{i-1-l}(k) s_{l} = 
A^{i} x(k)+\sum_{l=0}^{i} A^{i-1-l} B s_{l} + e_{i}(k)\label{p3}
\end{align}
 where the following error bounds hold for any $i=0...N$:
$\| e_i(k)\|^2 \le \omega_1(k) + \omega_2(k) \|x(k)\|^2 + \omega_3(k) \sum_{l=0}^i \|s_l\|^2$
with $\lim_{k \rightarrow \infty} \omega_j(k)=0$, $j=1,2,3$. 

(c) Finally, for any trajectory $x(k)$, $k \in \mathbb{N}$, of the closed loop
\eqref{mpc1} and \eqref{sysmpc}, 
there exists functions $\mu_0(k),...,\mu_{N-1}(k)$ such that
\begin{align}
\label{stabilizability}
\lim_{k \rightarrow \infty }\|A_{N}(k) x(k)+\sum_{l=0}^{N-1} 
B_{N-1-l}(k) \mu_{l}(k)\|=0.
\end{align}
\end{ass}


\begin{rem} \label{rk0} 
Assumption \ref{ass0} postulates predictor maps 
for the closed loop trajectory (input and state sequence) generated by \eqref{mpc1} and \eqref{sysmpc} (or \eqref{mpc3} and \eqref{sysmpc}). In particular, equation \eqref{predictor} and \eqref{p2}
in Assumption \ref{ass0}(a) ensure that we have 
linear time-varying and converging state predictor maps.
Equation \eqref{p4}  ensures a state property 
in the sense that the predictor maps commute like flow maps of (time-invariant) dynamical state-space models do.
Equation \eqref{p3}  ensures
that the predictor maps are able
to accurately predict  $x_1(x(k),p_1(k)),....,x_{N}(x(k),p_1(k))$ in \eqref{mpc1} \emph{along the closed loop trajectory}.
Notice that $e(k)$ converges to zero, if, for example, the state and input stays bounded.
Finally, equation \eqref{stabilizability}  
in Assumption \ref{ass0}(c) represents a stabilizability condition of the predictor maps along the closed loop trajectory.
Notice that if the state predictor maps are determined (learned) online, then Assumption \ref{ass0} does not imply that the knowledge of such state predictors implies  a model (system) identification in the sense that
neither the equation $(\hat A_i, \hat B_{l-i}) \approx (A^i,A^{l-i}B)$ must hold
nor that for \emph{every} initial data or \emph{every} input sequence the predictions are (asymptotically) accurate.
\end{rem}

The main goal of the next subsection is to show that the state of the closed loop \eqref{sysmpc}, \eqref{mpc1} converges to zero under the stated assumptions.
In summary, we aim for a receding horizon control scheme which guarantees that the state of a linear \emph{stabilizable} system converges to zero,  assuming that the closed loop state trajectory can be accurately predicted  as time goes to infinity. 
A characteristic property of the proposed scheme is that the objective (and potentially constraints) can be chosen independently from the system model (predictor maps) in the sense that no terminal cost or terminal constraint needs to be computed online based on some model information or data. 
This is a desirable property when controlling unknown systems.
Further, no stabilizing zero terminal constraint  and no controllability assumption
is required.
Instead, a time-varying terminal state weighting scheme is introduced to ensure convergence.

\subsection{Results}

We define the following auxiliary problems

\begin{align}
\begin{split}
     V_2(x,p_2) = \min ~&~\xi_N^\top Q_N \xi_N \label{mpc2}\\
    \text{s.t. } ~&~ \xi_{i + 1} = A_{i+1}(k)\xi_0+ \sum_{l=0}^iB_{i-l}(k) \nu_l, \\
    ~&~ \xi_0 = x, ~i=0...N-1  
\end{split}    
\end{align}
with $p_2=k$ and
\begin{align}
\begin{split}
     V_3(x,p_3) = \min ~&~ \sum\limits_{i = 0}^{N - 1}  \xi_i^{\top} Q \xi_i + \nu_i^{\top} R \nu_i   \label{mpc3}\\
    \text{s.t. } ~&~ \xi_{i + 1} = A_{i+1}(k) \xi_0+ \sum_{l=0}^iB_{i-l}(k) \nu_l,  \\
    ~&~ \xi_0 = x, ~i=0...N-1, \\
    ~&~ \xi_N = r 
\end{split}    
\end{align}
with $p_3^\top=[k, r^\top]$. A corresponding notation as for \eqref{mpc1} is used in \eqref{mpc2} and  \eqref{mpc3}.
~\\

\begin{lem} 
\label{mp}

\textcolor{black}{
Suppose Assumption \ref{ass1a}  holds true.
a) Let $\{x(k)\}_{k \in \mathbb{N}}$ and $\{r(k)\}_{k \in \mathbb{N}}$ be sequences
such that $\lim_{k \rightarrow \infty} r(k)=0$
and such that
problem \eqref{mpc3} is feasible for every time instant $k \in \mathbb{N}$
with $x=x(k),r=r(k)$.
Then the value  $V_3(x(k),p_3(k))$ of problem \eqref{mpc3} is given by
 $V_3(x(k),p_3(k))=x(k)^\top S_3(k) x(k) + x(k)^\top S_4(k) r(k) + r(k)^\top S_5(k) r(k) \ge 0$ for some 
 matrices $S_3(k),S_4(k),S_5(k)$ and
  the solution    of \eqref{mpc3}  is unique
 and linearly parameterized in $x(k),r(k)$ in the sense of
 $\xi_i(x(k),p_3(k))=K_{1,i}(k)x(k)+K_{2,i}(k)r(k)$ and 
 $\nu_i(x(k),p_3(k))=K_{3,i}(k)x(k)+K_{4,i}(k)r(k)$,
for all $i=0,...,N$.
 }
 b) Further, the value function $V_2(x(k),p_2(k))$ of problem \eqref{mpc2} is quadratic
 and positive semidefinite in $x$, i.e.
 $V_2(x(k),p_2(k))=x(k)^\top S_6(k) x(k) \ge 0$ for some 
 matrix $S_6(k)$, and there exists a
  solution  $\{\xi_i(x(k),p_2(k))\}_{i=0}^{N},\{\nu_i(x(k),p_2(k))\}_{i=0}^{N-1}$ 
  of \eqref{mpc3}  which is linearly parameterized in $x(k)$
  in the sense of a).
\end{lem}

\begin{ass}
\label{new}
\textcolor{black}{
 a)  Let $\{r(k)\}_{k \in \mathbb{N}}$ be a sequence which converges to zero.
 We assume that the
 solutions of \eqref{mpc3} along the 
 closed loop \eqref{mpc3}, \eqref{sysmpc}
 are uniformly bounded, i.e.
 $\xi_i(x(k),p_3(k))=K_{1,i}(k)x(k)+K_{2,i}(k)r(k)$,
 $\nu_i(x(k),p_3(k))=K_{3,i}(k)x(k)+K_{4,i}(k)r(k)$ 
 are bounded in the sense that there exists a bound $M>0$ such that for all $i=0,...,N$, $j=1...4$,  $k \in \mathbb{N}$ it holds  $\|K_{j,i}(k)\| \le M.$
 }
 \textcolor{black}{
b) We assume that the solutions
of \eqref{mpc1} along the closed loop \eqref{mpc1}, \eqref{sysmpc}
 are uniformly bounded, i.e.
 $x_i(x(k),p_1(k))=K_{5,i}(k)x(k)$,
 $u_i(x(k),p_1(k))=K_{6,i}(k)x(k)$ are bounded in the sense 
 that there exists a bound $M>0$ such that for all $i=0,...,N$, $j=5,6$,  $k \in \mathbb{N}$ it holds  $\|K_{j,i}(k)\| \le M.$
 }
\end{ass}

The main result of this subsection is Theorem \ref{MPC1vsMPC3}, which builds on the following two lemmas.

\begin{lem} 
\label{convergenceMPC3}
 Consider the closed loop \eqref{mpc3}, \eqref{sysmpc} and suppose 
 problem \eqref{mpc3} is feasible for every time instant $k \in \mathbb{N}$.
 Let Assumption \ref{ass1a} and \ref{ass0}\textcolor{black}{(a)(b)}  hold true and let $\{p_3(k)\}_{k \in \mathbb{N}}$ be a sequence  such that $\{r(k)\}_{k \in \mathbb{N}}$ converges to zero. 
 Then for any initial state $x(0)$, the state $x(k)$
 and the input $u(k)$
 of the closed loop
 converge to zero, i.e. $\lim_{k \rightarrow \infty} x(k) = 0$,
 $\lim_{k \rightarrow \infty} u(k) = 0$,
 \textcolor{black}{assuming that Assumption \ref{new} a) holds true}.
 \end{lem}

\begin{lem}
\label{lem1a}
Consider \eqref{mpc1} and suppose Assumption \ref{ass0}(a) holds true.
Further, suppose $\Gamma: \mathbb{R}^n \rightarrow \mathbb{R}$ is 
a function such that for all $x \in \mathbb{R}^n$, all $k \in \mathbb{N}$ 
 and all $\epsilon>0$
it holds that
$\Gamma(x) \ge c (\sum_{i=0}^{N-1} \|\xi_i(x,p_2)\|^2+\|\nu_i(x,p_2)\|^2)$ 
for some $c>0$,
where $p_2=k$
and $\{\nu_i(x,p_2)\}_{i=0}^{N-1}$, $\{\xi_i(x,p_2)\}_{i=0}^{N-1}$ is
some solution of \eqref{mpc2}.
Then there exists a $\rho>0$ such that solution $x_N(x,p_1)$, $p_1^\top=[k,\epsilon]$,
of
 \eqref{mpc1} satisfies for all $x \in \mathbb{R}^n$,
 all $k \in \mathbb{N}$ and all $\epsilon>0$
 \begin{align}
     \label{feasible}
     x_N(x,p_1)^\top Q_N x_N(x,p_1) \le V_2(x,p_2)+\epsilon \rho.
 \end{align}
 
 \end{lem}

\begin{thm} 
\label{MPC1vsMPC3}
Consider the closed loop \eqref{mpc1} and \eqref{sysmpc},
where  $\Gamma(x)=\alpha x^\top x$ for some $\alpha>0$ and
 suppose Assumption 
 \ref{ass1}, \ref{ass1a} and \ref{ass0}(a)-(c) hold true.
Let further $\{p_1(k)\}_{k \in \mathbb{N}}$ be a sequence such that  $\{\epsilon(k)\}_{
k \in\mathbb{N}}$, $\epsilon(k)>0$, converges to zero.
Then for any initial state $x(0)$, the state $x(k)$ of the closed loop converges to zero as $k$ goes to infinity,
\textcolor{black}{assuming that Assumption \ref{new} b) holds true}.
 \end{thm}

\textcolor{black}{Assumption \ref{new} is our main technical assumption that we impose in the proposed scheme. 
For example, if the predictor maps $P_N$ are controllable in the sense that
$\mathrm{rank}[B_0(k),...,B_{N-1}(k)]=n$ for all $k \in \mathbb{N}$
and also the limiting predictor has the same property, i.e. $\mathrm{rank}[\hat B_0,...,\hat B_{N-1}]=n$, then Assumption \ref{new} holds true (as discussed in the proof of Lemma \ref{mp}).
However, we admit that this assumption is in general difficult to verify. 
}
~\\~\\


\section{A Proximity-based Estimation  Scheme}
\label{estimationpart}

\subsection{Problem Setup}

Consider an output sequence (or some observed signal)  and an input sequence
\begin{align}
\label{signals}
\{y(k)\}_{k \in \mathbb{N}}, ~
\{v(k)\}_{k \in \mathbb{N}}
\end{align}
with $y(k) \in \mathbb{R}^{\bar p_y}$, $v(k) \in \mathbb{R}^{\bar q_v}$.
Let 
\begin{align}
\begin{split}
\label{liftedsignals}
x(k)&=\phi_{\mathrm{y}}(y(k),...,y(k-\bar N_y+1)) \in \mathbb{R}^{\bar n},\\ u(k)&=\phi_{\mathrm{v}}(v(k),...,v(k-\bar N_v+1)) \in \mathbb{R}^{\bar q}
\end{split}
\end{align}
and $\phi_{\mathrm{y}}: \mathbb{R}^{\bar p_y \bar N_y}
\rightarrow \mathbb{R}^{\bar n}$,
$\phi_{\mathrm{v}}: \mathbb{R}^{\bar q_v \bar N_v} \rightarrow \mathbb{R}^{\bar q}$
be some given basis (lifting) functions, e.g.
$\phi_{\mathrm{y}}(y_1(k),y_2(k),y_1(k-1),y_2(k-1))=[y_1(k), y_2(k), y_1(k-1), y_2(k-1), y_1(k) y_2(k), y_1(k)^2, y_2(k)^2]^T$, $\bar p=2, \bar n=6, \bar N_y=2$.
Notice that in principle one could also consider
cross-terms between input and output data, like $u(k)y(k)^3$, but such terms are not considered here for the sake of simplicity.
Consider, further, at time instant $k$ the optimization problem
\begin{align}
\label{sp1}
\begin{split}
     \theta^*(k)= & \mathrm{arg}~\min  ~~  c(e,k)+D(\theta,\tilde \theta(k-1))\\
    &~~~~~~\text{s.t. } ~~ s(k)-R(k) \theta =e \\
    \tilde \theta(k) = &(1-\lambda_k)\theta^*(k) + \lambda_k \tilde \theta(k-1)
\end{split}    
\end{align}
with
$\lambda_k \in [0,\lambda_{max})$, $\lambda_{max} \in (0,1)$,
where $\bar N \in \mathbb{N}$,
$c: \mathbb{R}^{\bar n \bar N} \times \mathbb{N} \to \mathbb{R}$
and 
$
D(x,y) = g(x) - g(y) - (x - y)^{\top} \nabla_y g(y)
$
defines the Bregman distance induced by a function $g: \mathbb{R}^{\bar n \bar N} \to \mathbb{R}$. 
The vector $s(k)$ is defined as
\begin{align}
\label{sigpred2}
 s(k)=\begin{bmatrix} x(k)^\top \ldots x(k-\bar N +1)^\top  \end{bmatrix}^\top
\end{align}
and the matrix $R(k)$ is defined as 
\begin{align}
\label{sigpred}
\begin{split}
R(k) &=  \begin{bmatrix}
x(k - 1)^T \otimes I &  u(k-1)^T \otimes I \\
\vdots & \vdots  \\
x(k - \bar N)^T \otimes I &  u(k-\bar N)^T \otimes I \\
\end{bmatrix}.  
\end{split}
\end{align}

 Decision variables are the parameter vector $\theta \in \mathbb{R}^{\bar n (\bar n + \bar q)}$ and $e$. We refer to $\theta$ when using $\mathrm{arg}\min$ 
 since the (slack) variables $e$ can be eliminated and have been introduced just for the notational convenience. 
Also we define in the following $y(k)=0,v(k)=0,v_k=0$ etc. whenever $k<0$.
We impose now the following assumptions. 
\begin{ass}
\label{zerovalue}
The objective function $c$ in \eqref{sp1} is  continuously differentiable and strictly convex in the first argument
and it satisfies for all $k$ and $e \not = 0$: $c(e,k)>c(0,k)$. 
Further, the function $g$, which defines the Bregman distance $D$, is  continuously differentiable and strictly convex.
\end{ass}

\begin{ass}
\label{exo}
For the given sequences in \eqref{signals}
and given $x(k)=\phi_{\mathrm{y}}(y(k),...,y(k-\bar N_y+1))$ and $u(k)=\phi_{\mathrm{v}}(v(k),...,v(k-\bar N_v+1))$
in \eqref{liftedsignals},
there exist matrices $A,B$ and
$x_0 \in \mathbb{R}^{\bar n}$ 
that satisfy
\begin{align}
\label{eq:exo}
\begin{split}
x(k + 1) &= A x(k) + B u(k), \quad x(0) = x_0.
\end{split}
\end{align}
\end{ass}

\begin{rem}
a) Notice that $s(k)=R(k)\theta$
with $\theta^\top = [\mathrm{vec}(A)^\top,~\mathrm{vec}(B)^\top]$,
where $\mathrm{vec}(A)$ corresponds to the (column-wise) vectorization of a matrix $A$,
is the linear
system of equations $x(j)=Ax(j-1)+Bu(j-1)$, $j=k...k-\bar N+1$.
b) If $c(e,k)=\|e\|^2,g(x)=\|x\|^2$, then \eqref{sigpred} reduces to a  least squares parameter estimation problem, where a closed form solution to it is known.
The motivation for a general convex cost is its flexibility
in tuning the estimator. Similarly as in a recently proposed state estimation scheme based on proximal minimization \cite{Gha-Gha-Ebe-20},
specifying different $c,D$ allows to take into account various aspects like  outliers in the data, sparsity in the parameters or cost-biased objectives \cite{Bit-06}.
\end{rem}
\begin{rem}
\label{rk4}
Assumption \ref{exo} 
imposes that the given (lifted) signal $\{x(k)\}_{k \in \mathbb{N}}$ 
can be reproduced by some linear time-invariant system
that is driven by the given (lifted) input sequence $\{u(k)\}_{k \in \mathbb{N}}$.
Notice that reproducing a given signal by \eqref{eq:exo} does not imply that the signal $x(k)$ itself originates from \eqref{eq:exo} nor by a linear time-invariant system at all.
For example, a given \emph{(single)} trajectory of a nonlinear system or even all
trajectories of a large class of nonlinear systems can be reproduced by or embedded into high dimensional linear (not necessarily controllable) systems using for example Carleman or Koopman lifting
techniques. 
Hence,  \eqref{eq:exo} represents a signal model of the actual closed-loop trajectory, rather than a system model of all possible trajectories of the plant.
A signal model is therefore a parsimonious modeling approach in the sense that it
aims to predict nothing more than the closed-loop trajectory.

\end{rem}

The main goal of the next subsection is to show
that the (parameter) estimates $\tilde \theta(k)$ ($\theta^*(k)$) obtained from \eqref{sp1} 
converge and that the estimates can be used to define $i$th step-ahead signal predictor maps $v_i=P_i(k,v_0,u_0,...,u_{i-1})$ which have the properties as described in  Assumption \ref{ass0} 
for the signal model \eqref{eq:exo}
and for the given data \eqref{liftedsignals}.
The convex combination  in \eqref{sp1} is introduced to resolve the loss
of stabilizability problem (see next subsection).
In summary, we therefore aim for an estimation scheme to obtain
asymptotically accurate predictor maps for a given input and output sequence that can be embedded in a potentially high-dimensional linear signal model.
An important property of the proposed scheme is that no system identification is carried out and no persistency of excitation condition is needed. The online computational burden is again rather low since, 
in its simplest form, the problem boils down to a least-squares regression problem.

\subsection{Results}

We first prove a lemma which is a key step for the
convergence of the estimation scheme. It provides a convergence result for a proximal minimization scheme with a time-varying objective function which has at least one common (time-invariant) minimizer.

\begin{lem}
\label{timevaryingmin}
Let $f: \mathbb{R}^n \times \mathbb{N} \to \mathbb{R}$ be convex and 
continuous differentiable in the first argument.  Suppose the set of minimizers 
$\mathcal{X}_k = \lbrace x^*_k \in \mathbb{R}^n :~ f(x,k) \ge f(x^*_k,k):=0  \; \forall x \in \mathbb{R}^n \rbrace$
of $f$ at any time instant $k$ is nonempty and also their intersections
\begin{align}
\mathcal{X} = \bigcap_{k = 0}^{\infty} \mathcal{X}_k \neq \emptyset, \label{eq:nonempty2}
\end{align}
i.e. there exists a common (time-invariant) minimizer $x^* \in \mathcal{X}$ which minimizes $f$ for
any $k$ with a common minimum value zero. 
Let further $g:\mathbb{R}^n \to \mathbb{R}$ be strictly convex, continuous differentiable, $\lambda_{max} \in  [0,1)$, and let
$D$ denotes the Bregman distance induced by $g$. 
Assume that $D$ is convex in the second argument,
then the proximal minimization iterations $x^*_{k},\tilde x_{k}$ given by
\begin{align}
\begin{split}
x^*_{k + 1} &= \mathrm{arg} \min\limits_{x} f(x,k) +  D(x, \tilde x_k)\\
\tilde x_{k+1}   &= (1-\lambda_{k+1}) x^*_{k + 1} + \lambda_{k+1} \tilde x_k
\end{split}
\label{eq:prox_iter2}
\end{align} 
with $\lambda_{k+1} \in [0,\lambda_{max}]$,
converge to a point in $\mathcal{X}$, i.e.  $\lim_{k \rightarrow \infty} x^*_k
=\lim_{k \rightarrow \infty} \tilde x_k=\tilde x^0 \in \mathcal{X}$.
\end{lem}

Notice that the Bregman distance is in general not convex in the second
argument, but there are important cases, such as $g(x)=x^T Q x$, $Q>0$, where this holds \cite{Bau-Bor-01}.
Notice further that in a classical proximal minimization scheme $\tilde x_{k+1}=x^*_{k+1}$ ($\lambda_{k+1}=0$). Here, we pick $\tilde x_{k+1}=(1-\lambda_{k+1}) x^*_{k + 1} + \lambda_{k+1} \tilde x_k$ instead of $x^*_{k + 1}$ as the next iterate, because with appropriately chosen $\lambda_{k+1}$'s,
the loss of stabilizability problem
in the estimation scheme \eqref{sp1}
can be avoided and it can be guaranteed  that every estimated signal model is controllable, if the initial model is controllable,
as shown next.

\begin{lem}
\label{staycontrollable}
Consider $A_c,A_u \in \mathbb{R}^{n \times n}$,
$B_c,B_u \in \mathbb{R}^{n \times q}$, $q \le n$, and
let $(A_c,B_c)$ be controllable and $(A_u,B_u)$ be not controllable.
Then for any $\lambda_{max} \in (0,1)$, there exists a $\lambda \in (0,\lambda_{max})$
such that $(A(\lambda),B(\lambda))$
with $A(\lambda)=(1-\lambda)A_u+\lambda A_c$, $B(\lambda)=(1-\lambda)B_u+\lambda B_c)$ is controllable.
In particular, take some $\lambda_j$'s with $0<\lambda_1<...<\lambda_{2n^2+1}<\lambda_{max}$,
then there exists an $i \in \{1,...,2n^2+1\}$ such that $(A(\lambda_i),B(\lambda_i))$
is controllable.
\end{lem}

The next theorem is the main result of this subsection.

\begin{thm}
\label{xxx}
Consider the sequences \eqref{signals} with some given basis functions \eqref{liftedsignals}
and consider the 
optimization problem \eqref{sp1}.
Suppose Assumption \ref{zerovalue} and \ref{exo} hold true
and assume that $D$ is convex in the second argument.
Then the following statements hold true.

(i) The solution sequence $\{\tilde \theta(k)\}_{k \in \mathbb{N}}$ converges, i.e.
$\lim_{k \rightarrow \infty}\tilde \theta(k)=\theta^*$. 

(ii) If one defines $\tilde \theta(k)^\top = [\mathrm{vec}(A(k))^\top,~\mathrm{vec}(B(k))^\top]
$ and predictor maps \eqref{predictor} according to 
\begin{align}
\label{predictormaps}
\begin{split}
 A_i(k)=& A(k)^i \\
 B_i(k)=& A(k)^i B(k), 
\end{split}  
\end{align}
$i=0...\bar N$, 
then the predictor fulfills the properties \eqref{p2}, \eqref{p4} and \eqref{p3} in Assumption \ref{ass0}(a)(b) with respect to 
the signal model \eqref{eq:exo} and the sequences \eqref{liftedsignals}.

(iii) 
In addition, if the initialization $(A(0),B(0))$,
$\tilde \theta(0)^\top = [\mathrm{vec}(A(0))^\top,~\mathrm{vec}(B(0))^\top]$,
of \eqref{sp1} is controllable, then there exists a sequence $\{\lambda_k\}_{k \in \mathbb{N}}$, $\lambda_{max} \in (0,1)$, (constructed for example according to Lemma \ref{staycontrollable}) such that for any $k \in \mathbb{N}$ the pair $(A(k),B(k))$
is controllable and hence also 
\textcolor{black}{
Assumption} \ref{ass0}(c)  is fullfilled.
\end{thm}


\section{The Overall Scheme}
\label{overallpart}

In Section \ref{rhcpart}, Theorem \ref{MPC1vsMPC3}, we have established a control scheme
which drives the state of the linear system \eqref{sysmpc} to zero assuming that the system is stabilizable
and that state measurements as well as
asymptotically accurate predictor maps are available.
In Section \ref{estimationpart}, Theorem \ref{xxx}, we have established an estimation scheme
which delivers asymptotically accurate predictor maps
for any lifted signals \eqref{liftedsignals} assuming that these signals can be embedded in a linear signal model  of the form \eqref{eq:exo}.
Utilizing the estimation scheme \eqref{sp1} in the control scheme \eqref{mpc1} means now that the signal model \eqref{eq:exo} replaces
the system model \eqref{sysmpc}
 and the predictor maps in \eqref{predictormaps}
 are used to define the predictor \eqref{predictor}.
However, it needs to be clarified how the output and input sequence $\{y(k)\}_{k \in \mathbb{N}}$, $\{v(k)\}_{k \in \mathbb{N}}$ of \eqref{sys}
under Assumption \ref{ass1ground}
can be related to a signal model of the form \eqref{eq:exo} such that 
$(A,B)$ is stabilizable and such that the (observable part of the) state $x(k)$ is available.
This issue is addressed in the next lemma.

\begin{lem}
\label{finalstep}
Consider an arbitrary output and input sequence $\{y(k)\}_{k \in \mathbb{N}}$,
$\{v(k)\}_{k \in \mathbb{N}}$ of system \eqref{sys} and suppose Assumption \ref{ass1ground} holds true. Let
\begin{align}
\label{arx}
\begin{split}
 x(k)&=\phi_{\mathrm{y}}(y(k),...,y(k-m+1))=
 \begin{bmatrix}
    y(k)^\top,...,y(k-m+1)^\top
 \end{bmatrix}^\top,~\\
 u(k)&=\phi_{\mathrm{v}}(v(k),...,v(k-m+1))=
  \begin{bmatrix}
    v(k)^\top,...,v(k-m+1)^\top
 \end{bmatrix}^\top
\end{split} 
\end{align}
with $m \ge n$.
Then the sequences $\{u(k)\}_{k \in \mathbb{N}}$,
$\{x(k)\}_{k \in \mathbb{N}}$ satisfy Assumption \ref{exo}
with a stabilizable pair of matrices $(A,B)$.
In addition, if the sequences $\{u(k)\}_{k \in \mathbb{N}}$,
$\{x(k)\}_{k \in \mathbb{N}}$ converge to zero when $k$ goes to infinity,
then so do the sequences $\{v(k)\}_{k \in \mathbb{N}}$,
$\{y(k)\}_{k \in \mathbb{N}}$.
\end{lem}

\begin{rem}
\label{intchain}
Notice that the lifted input vector $u(k)$ in \eqref{arx}
contain past values of the actual input vector $v(k)$. 
In order to obtain a state space model with input $v(k)$, one can
just add state variables to the signal model. 
In more detail, define an integrator chain
dynamics of the form $\zeta_1(k+1)=\zeta_2(k),...,
\zeta_{m-2}(k+1)=\zeta_{m}(k), \zeta_{m-1}(k+1)=v(k)$,
hence $\zeta_1(k)$ corresponds to $v(k-m)$ etc.
This augmentation does not effect the stabilizability property,
since the states of the integrator chain converge to zero, if $v(k)$ converges
to zero.
This state augmentation in the signal model 
leads to matrices with at least the size $A \in \mathbb{R}^{mp+(m-1)q \times mp+(m-1)q}$,
$B \in \mathbb{R}^{mp+(m-1)q \times q}$
and needs to be taken into account when implementing the receding horizon scheme.
\end{rem}

We are now ready to close the loop.
By Lemma \ref{finalstep}, we known that the output and input sequences
$\{y(k)\}_{k \in \mathbb{N}}$, $\{v(k)\}_{k \in \mathbb{N}}$ of
system \eqref{sys} satisfy Assumption \ref{exo}
and Assumption \ref{ass1} w.r.t. the signal model \eqref{eq:exo} 
(equation \eqref{hsm}). Assumption \ref{ass1a}
and Assumption \ref{zerovalue} can be satisfied by setting up
the optimization problem accordingly.
By Theorem \ref{xxx}, 
 Assumption \ref{ass0}  holds. 
Hence all assumptions are satisfied and thus the receding horizon scheme  guarantees, by Theorem \ref{MPC1vsMPC3} 
with $\Gamma(x)=\alpha x^Tx,\alpha>0$, a sequence
$\epsilon(k) \rightarrow 0$, and Assumption \ref{new} b), 
together with Lemma \ref{finalstep}
that the state and the input of \eqref{sys} converges to zero.
These arguments lead to the next theorem.

\begin{thm} 
\label{mainthm}
Consider the closed loop system
consisting of the system \eqref{sys},
the receding horizon control scheme \eqref{mpc1}
and the proximity-based estimation scheme \eqref{sp1}.
Define $x(k)$, $u(k)$ ($\phi_{\mathrm{y}},\phi_{\mathrm{v}}$) according to equation \eqref{arx}
and set up the predictor scheme \eqref{sp1} according
to Assumption \ref{zerovalue}.
Further, set up the receding horizon scheme \eqref{mpc1} according to 
Assumption \ref{ass1a} with $n=m$, $\alpha>0,\Gamma(x)=\alpha x^T x$
and a sequence $\{\epsilon(k)\}_{k \in \mathbb{N}}$, $\epsilon(k)>0$
that converges to zero.
Then, under Assumption \ref{ass1ground}, 
for any initial state $z(0)$, the state $z(k)$ of the closed loop and the input $v(k)$ of the closed 
loop converges to zero as $k$ goes to infinity,
\textcolor{black}{assuming that Assumption \ref{new} b) holds true}.
 \end{thm}

\tikzset{
input/.style={coordinate},
output/.style={coordinate}
}

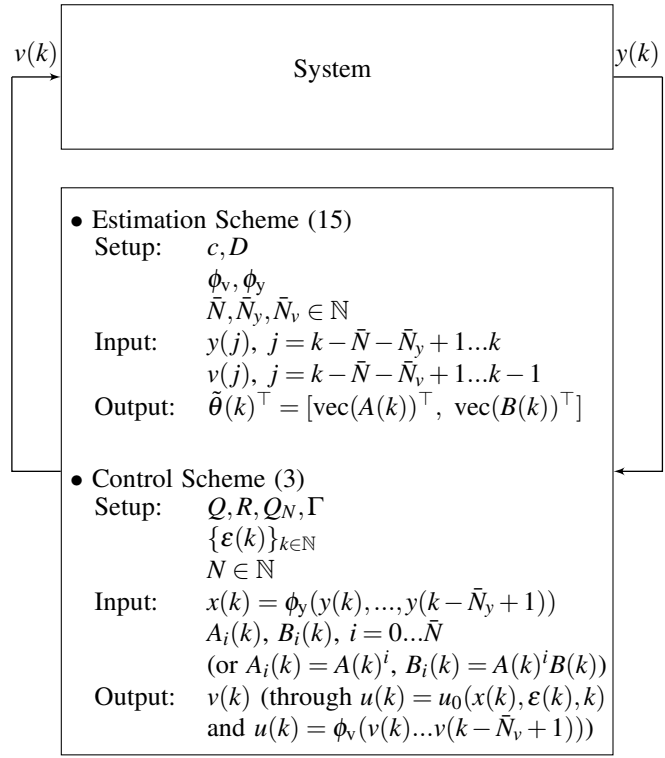
\begin{figure}[ht!]
\begin{center}
\begin{tikzpicture}[auto, node distance=0.65cm,>=latex']
    \setlength{\abovedisplayskip}{0pt}
    \setlength{\belowdisplayskip}{0pt}

		\node [input, name=input] {};
		
    \node [rectangle, draw, right= of input, text width=7.1cm] (system) at (input.east) {
        \begin{minipage}{\textwidth}
				 \begin{center}
				~\\~\\
				 System
				~\\~\\~\\
				 \end{center}
        \end{minipage}
    };
		
    \node [rectangle, draw, below=0.5cm, text width=7.1cm] (controller) at (system.south) {
        \begin{minipage}{\textwidth}~\\
				 $\bullet$ Estimation Scheme \eqref{sp1} \\ 
				 \begin{tabular}{ll}
				 ~Setup: & $c,D$   \\
				             & $\phi_{\mathrm{v}},\phi_{\mathrm{y}}$  \\
										  & $\bar N, \bar N_y, \bar N_v \in \mathbb{N}$ \\
         ~Input:      & $y(j)$, $j=k-\bar N-\bar N_y+1...k$  \\
         & $v(j)$, $j=k-\bar N- \bar N_v +1...k-1$		\\
         ~Output:    & $    \tilde \theta(k)^\top = [\mathrm{vec}(A(k))^\top,~\mathrm{vec}(B(k))^\top]$  
				 \end{tabular}
~\\~\\
				 $\bullet$  Control Scheme \eqref{mpc1} \\ 
				 \begin{tabular}{ll}
				 ~Setup: & $Q,R,Q_N,\Gamma$ \\
				             & $\{\epsilon(k)\}_{k \in \mathbb{N}}$ \\
				             & $N \in  \mathbb{N}$   \\
         ~Input:      & $x(k)=\phi_{\mathrm{y}}(y(k),...,y(k-\bar N_y+1))$  \\
         			&  $A_i(k)$, $B_i(k)$, $i=0...\bar N$  \\
         			& (or $A_i(k)= A(k)^i$, $B_i(k)= A(k)^i B(k)$) \\		    	   
				    ~Output:    	    
				    & $v(k)$ (through $u(k)=u_0(x(k),\epsilon(k),k)$\\ & and $u(k)=\phi_{\mathrm{v}}(v(k)...v(k-\bar N_v+1)) $)
				 \end{tabular}

        \end{minipage}
    };
		
		\node [output, right =of system, name=output] {};

    \draw [->, line width=0.5pt] (input) -- node[name=u] {$v(k)$} (system);
		\draw [-, line width=0.5pt] (system) -- node[name=y] {$y(k)$} (output);
		\draw [->, line width=0.5pt] (output) |- (controller);
		\draw [-, line width=0.5pt] (controller) -| (input);
\end{tikzpicture}
\end{center}
\caption{The overall scheme}
\label{eq:overall_scheme}
\end{figure}

\textcolor{black}{We conclude with some remarks.
Firstly, if we postulate that the limiting signal model $\hat A, \hat B$ is controllable, then we can drop Assumption \ref{new} b) in Theorem \ref{mainthm}}.
This is for example the case, when the signal model is controllable
and when the estimates are unique (persistently excitation).
Notice further that the state extension
in Remark \ref{intchain} has to be taken into account when implementing the overall scheme.
Moreover, if $\phi_\mathrm{v},\phi_\mathrm{y}$
are nonlinear functions (notice that in \eqref{arx} these are linear functions), then this requires further considerations (e.g. one has to be able to extract the input $v$ from $\phi_\mathrm{v}$), which is out of the  scope of this work.
Secondly, notice that the convergence result of Theorem \ref{mainthm}
also holds for certain classes of nonlinear systems. As already mention
in Remark \ref{rk4}, certain (trajectories of) nonlinear
systems can be embedded into high-dimensional linear (signal) models, e.g. 
$z_1(k+1)=z_1(k)+z_2(k)^2+u(k), z_2(k+1)= 0.5 z_2(k)$,
which can be written as a linear system with an additional state variable $z_3(k)=z_2(k)^2$, see \cite{Carleman32,Koo-31,Korda17,Mue-16}.
Thirdly, if a model of \eqref{sys} is known, then 
it follows from the proofs of Theorem \ref{MPC1vsMPC3} and \ref{mainthm} 
that global asymptotic stability (instead of global convergence) can be guaranteed.
Moreover, in applications often models are available for  at least some parts of the system. The proposed scheme can eventually be applied to such situations,
for example, if we have two subsystem $x_1(k+1)=F_1x_1(k)+F_2x_2(k)+G_1u(k),x_2(k+1)=F_3x_1(k)+F_4x_2(k)+G_2u(k),
y(k)=x_1(k)$,
where $F_3,F_4,G_2$ is unknown, than one can consider the second
subsystem as the unknown system with input $u(k),x_1(k)$ and adjust the
receding horizon control scheme accordingly.
It is also possible to model unknown disturbances or reference  signals 
by a socalled unknown  exosystem and to tread this system as a signal  model
\cite{Mue-16}.
Finally, a priori knowledge about the system model can be taken into account
by initializing the estimator appropriately or by including parameter constraints in the estimator.

\section{Simulation Results}
\label{secsim}
In the following, we show simulation results of the proposed
overall scheme for a linear and a nonlinear system.


\subsection{Linear system}
We consider the system
\begin{align}
\begin{split}
    x(k + 1) &= \begin{bmatrix}
    0 & 1 & 0.1 \\
    0 & 1.02 & 0 \\
    0 & 0 & 0.92
    \end{bmatrix} x(k) + \begin{bmatrix}
    0 \\
    1 \\
    0
    \end{bmatrix} u(k) \label{eq:lin_Model2}\\
    y(k) &= \begin{bmatrix}
    1 & 0 & 1
    \end{bmatrix} x(k).
\end{split}
\end{align}
The system is unstable, stabilizable and observable.
For the estimation scheme, we choose
$c(e)=\|e\|^2$,
$g(x)=\|x\|^2$ ($D(x,y)=\|x-y\|^2$),
$\bar N=8$,
$\bar N_y=4$,
$\bar N_v=4$,
$\phi_\mathrm{v}(v(k),...,v(k-3))=[v(k),...,v(k-3)]^\top$,
$\phi_\mathrm{y}(y(k),...,y(k-3))=[y(k),...,y(k-3)]^\top$.
Further, we initialize the estimator with a controllable signal model.
In this example, we assumed that we know that the system order is
at most four, i.e. $m=4$, hence we have chosen $\bar N_y=4=\bar N_v$.
For the control scheme, we choose
$Q=100 I$,
$R=10000 I$,
$Q_{N}=100 I$,
$\Gamma(x)=x^T x$,
$\epsilon(k)=\frac{1}{1+1000k}$,
$N=20$.
Some simulation results 
for the initial condition to be $x(0) = \begin{bmatrix} 0.1 & 0.1 & -10 \end{bmatrix}^{\top}$
are shown in 
the Figures \ref{subfig1} to \ref{subfig:gamma}.
It can be verified that the state converges to zero
(see Figure \ref{subfig4} for evolution of the state $x_1(k)$).

\begin{figure}[ht]
\centering
%
%
\begin{tikzpicture}

\begin{axis}[%
width=0.856\columnwidth,
height=2.5cm,
at={(0\columnwidth,0cm)},
scale only axis,
xmin=0,
xmax=40,
xlabel style={font=\color{white!15!black}},
xlabel={$k$},
ymin=-50,
ymax=35,
ylabel style={font=\color{white!15!black}},
ylabel={$y(k)$, $\hat y(k)$},
axis background/.style={fill=white},
xmajorgrids,
ymajorgrids,
ylabel style={yshift=-6pt}
]
\addplot[const plot, color=black, mark=x, mark options={solid, black}, forget plot] table[row sep=crcr] {%
0	-9.9\\
1	0\\
2	-21.1746638601147\\
3	0.997928957068858\\
4	18.9913177815077\\
5	27.8121854227279\\
6	-6.48430087460724\\
7	-16.7540526963875\\
8	-48.2483830493242\\
9	-40.3383973058427\\
10	31.0084954565087\\
11	22.9330205837194\\
12	16.8748305877275\\
13	10.2859234145828\\
14	7.60084959016017\\
15	7.02738436651966\\
16	6.49976299320551\\
17	5.40824550509399\\
18	4.49305311409806\\
19	3.73296818736471\\
20	3.10622948973561\\
21	2.58876622644501\\
22	2.16153825401701\\
23	1.80830397704626\\
24	1.51591860043003\\
25	1.27371616376073\\
26	1.07281681693003\\
27	0.905901456687476\\
28	0.766943172400388\\
29	0.651064656638543\\
30	0.554218785442641\\
31	0.473122520640297\\
32	0.405082213027904\\
33	0.347877818502019\\
34	0.299677821979591\\
35	0.258970117980181\\
36	0.224505367256338\\
37	0.195250540156131\\
38	0.170350762045022\\
39	0.149097955322346\\
40	0.130905061455309\\
};
\addplot[const plot, color=black, mark=o, mark options={solid, black}, forget plot] table[row sep=crcr] {%
0	-9.9\\
1	-10.1\\
2	-5.47293283091494\\
3	2.37390316586362\\
4	34.5915839195456\\
5	19.5461156948019\\
6	-3.99327224796716\\
7	-10.6122628749706\\
8	-39.6690568469463\\
9	-38.8544862274442\\
10	34.0636774764554\\
11	24.7446283511858\\
12	17.1438132542074\\
13	10.3154059764415\\
14	7.62806367921073\\
15	7.02843166385351\\
16	6.49994939305619\\
17	5.40830912958854\\
18	4.49312148811215\\
19	3.73303289266111\\
20	3.10628721927424\\
21	2.58882106455952\\
22	2.16158701117864\\
23	1.80835239704419\\
24	1.51596296879757\\
25	1.27375594837032\\
26	1.07285498168087\\
27	0.905939220521367\\
28	0.766977657554791\\
29	0.651095060980283\\
30	0.55424560361744\\
31	0.473146219921902\\
32	0.405103184469051\\
33	0.347896395153884\\
34	0.29969428652688\\
35	0.258984712708183\\
36	0.224518302845613\\
37	0.195262001283202\\
38	0.170360911511354\\
39	0.149106937936534\\
40	0.130913007460455\\
};

\end{axis}
\end{tikzpicture}%
\caption{Evolution of the true system output $y(k)$ $(-\circ-)$ ~ and the estimated system output $\hat y(k)$ $(-\times-)$}
\label{subfig1}
\end{figure}
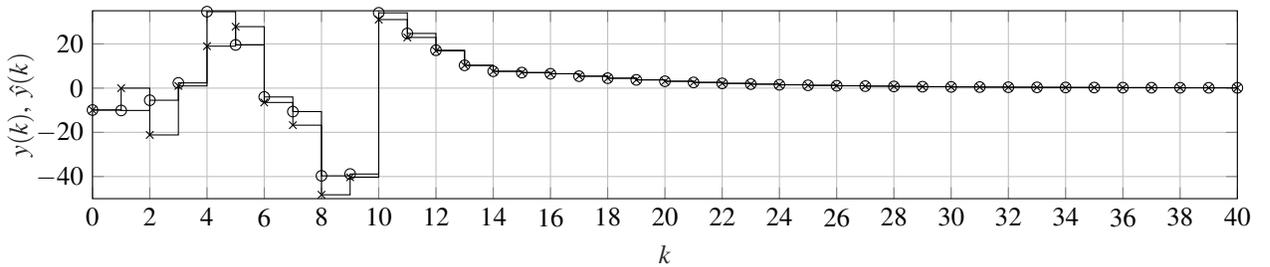
\begin{figure}[ht]
\centering
\input{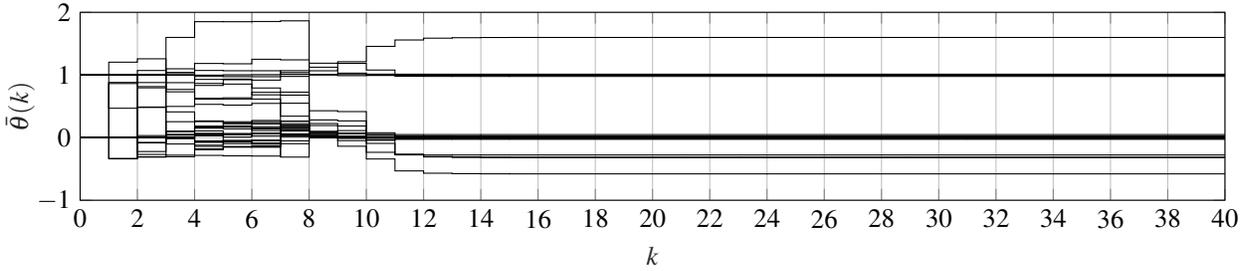}
\caption{Evolution of the estimates $\bar \theta(k)$}
\label{subfig2}
\end{figure}
\begin{figure}[ht]
\centering
%
%
\begin{tikzpicture}

\begin{axis}[%
width=0.856\columnwidth,
height=2.5cm,
at={(0\columnwidth,0cm)},
scale only axis,
xmin=0,
xmax=40,
xlabel style={font=\color{white!15!black}},
xlabel={$k$},
ymin=-30,
ymax=75,
ylabel style={font=\color{white!15!black}},
ylabel={$u(k)$},
axis background/.style={fill=white},
xmajorgrids,
ymajorgrids,
ylabel style={yshift=-6pt}
]
\addplot[const plot, color=black, forget plot] table[row sep=crcr] {%
0	3.80906716908506\\
1	7.01789465339685\\
2	31.3068746903647\\
3	-16.5315616631347\\
4	-24.6610310758651\\
5	-7.2113883357081\\
6	-29.4630308158472\\
7	1.03894822317989\\
8	73.1717701778481\\
9	-10.481927265351\\
10	-8.53878388731028\\
11	-7.57891366829344\\
12	-3.26867013216761\\
13	-1.09721142711677\\
14	-0.986467591195703\\
15	-1.51366260348038\\
16	-1.29201530804764\\
17	-1.0971195904744\\
18	-0.928801411280333\\
19	-0.788795372737219\\
20	-0.671477670418118\\
21	-0.57353617444439\\
22	-0.491460710666969\\
23	-0.4223981755243\\
24	-0.364258229720266\\
25	-0.315224433331465\\
26	-0.273783794112277\\
27	-0.238589320722055\\
28	-0.208649155899184\\
29	-0.183059869299284\\
30	-0.161111487500174\\
31	-0.142225938441522\\
32	-0.125923755132098\\
33	-0.111806080663089\\
34	-0.09954051554148\\
35	-0.0888495261300902\\
36	-0.0795009596274177\\
37	-0.0713002512485608\\
38	-0.0640840039228416\\
39	-0.0577146955422094\\
40	-0.0520763160253373\\
};
\end{axis}
\end{tikzpicture}%
\caption{Evolution of the control input $u(k)$}
\label{subfig3}
\end{figure}
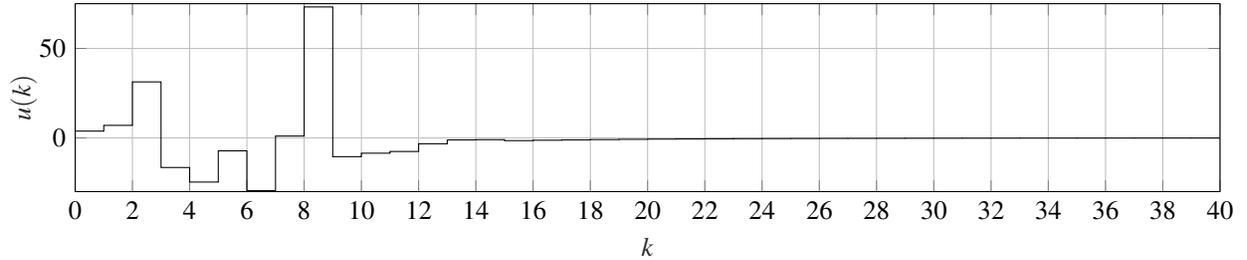
\begin{figure}[ht]
\centering
%
%
\begin{tikzpicture}

\begin{axis}[%
width=0.856\columnwidth,
height=2.5cm,
at={(0\columnwidth,0cm)},
scale only axis,
xmin=0,
xmax=40,
xlabel style={font=\color{white!15!black}},
xlabel={$k$},
ymin=-35,
ymax=42,
ylabel style={font=\color{white!15!black}},
ylabel={$x_1(k)$},
axis background/.style={fill=white},
xmajorgrids,
ymajorgrids,
ylabel style={yshift=-6pt}
]
\addplot[const plot, color=black, forget plot] table[row sep=crcr] {%
0	0.1\\
1	-0.9\\
2	2.99106716908506\\
3	10.1607831658636\\
4	41.7555135195456\\
5	26.1369309268019\\
6	2.07027776547284\\
7	-5.0337968626058\\
8	-34.5368681155707\\
9	-34.1328725945786\\
10	38.4075620186917\\
11	28.7410021300432\\
12	20.8204771307562\\
13	13.6979367428664\\
14	10.7399919843217\\
15	9.89140570455556\\
16	9.13388551050208\\
17	7.83153035763876\\
18	6.72248501791835\\
19	5.78404734008282\\
20	4.9932205109022\\
21	4.32479969285725\\
22	3.75868734921255\\
23	3.27768470803539\\
24	2.86774869490947\\
25	2.51739881639327\\
26	2.21700642026199\\
27	1.95855854401599\\
28	1.73538743516985\\
29	1.54203205638613\\
30	1.37390763939082\\
31	1.22723529283341\\
32	1.09886513154764\\
33	0.986157386466189\\
34	0.8868943985342\\
35	0.799208815754918\\
36	0.721524477648609\\
37	0.652507682101958\\
38	0.59102693786461\\
39	0.536119682181529\\
40	0.486964732165851\\
41	0.442859458709408\\
};
\end{axis}
\end{tikzpicture}%
\caption{Evolution of the system state $x_1(k)$}
\label{subfig4}
\end{figure}
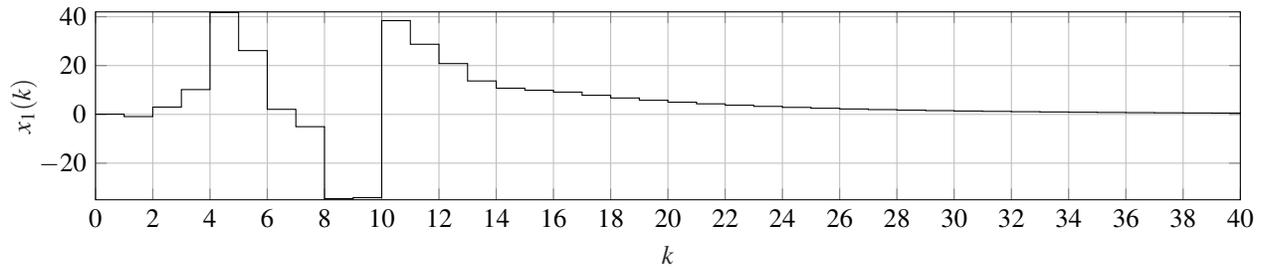 
\begin{figure}[ht]
\centering
%
%
\begin{tikzpicture}

\begin{axis}[%
width=0.856\columnwidth,
height=2.5cm,
at={(0\columnwidth,0cm)},
scale only axis,
xmin=0,
xmax=40,
xlabel style={font=\color{white!15!black}},
xlabel={$k$},
ymin=-1,
ymax=10000000,
xmajorgrids,
ymajorgrids,
ylabel style={font=\color{white!15!black}},
ylabel={$\frac{\Gamma(x(k))}{\epsilon(k)}$},
axis background/.style={fill=white},
ylabel style={yshift=-6pt}
]
\addplot[const plot, color=black, forget plot] table[row sep=crcr] {%
0	112.518992698602\\
1	26641.7636445426\\
2	256044.500177651\\
3	463019.92897298\\
4	1281497.94575593\\
5	1276387.59648598\\
6	1880198.88068475\\
7	1842472.21675345\\
8	6654314.39315011\\
9	7817913.6428199\\
10	9903042.14710059\\
11	5610869.39477787\\
12	4464413.8336962\\
13	2917201.98481666\\
14	1518212.0265577\\
15	769070.831610618\\
16	418143.970442914\\
17	313408.410267741\\
18	260824.042218636\\
19	205899.907974196\\
20	149991.469611599\\
21	108992.976582264\\
22	79192.7886758265\\
23	57622.8136322392\\
24	42006.3904111696\\
25	30696.4373093083\\
26	22495.4305071536\\
27	16539.8179270514\\
28	12206.3925432557\\
29	9045.00886533626\\
30	6731.53068618075\\
31	5032.70929425865\\
32	3780.79976804714\\
33	2854.64417954163\\
34	2166.74379960283\\
35	1653.63920126949\\
36	1269.18351218782\\
37	979.742898362684\\
38	760.741745179782\\
39	594.172654256258\\
40	466.800159313449\\
};
\end{axis}
\end{tikzpicture}%
\caption{Evolution of  $\Gamma(x(k))/\epsilon(k)$}
\label{subfig:gamma}
\end{figure}
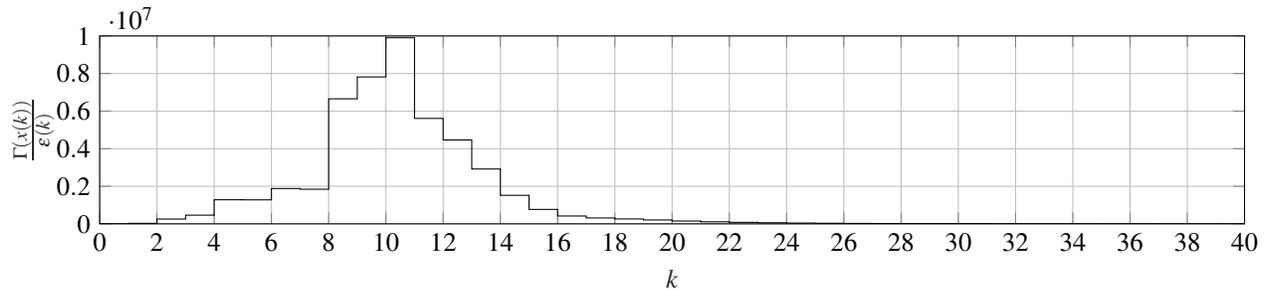 

\subsection{Nonlinear system}
\begin{figure}[t]
\centering
%
%
\begin{tikzpicture}

\begin{axis}[%
width=0.856\columnwidth,
height=2.5cm,
at={(0\columnwidth,0cm)},
scale only axis,
xmin=0,
xmax=60,
xlabel style={font=\color{white!15!black}},
xlabel={$k$},
ymin=-18,
ymax=16,
ylabel style={font=\color{white!15!black}},
ylabel={$x_3(k)$},
axis background/.style={fill=white},
xmajorgrids,
ymajorgrids,
ylabel style={yshift=-6pt}
]
\addplot[const plot, color=black, forget plot] table[row sep=crcr] {%
0	1\\
1	0.814869682199767\\
2	-1.14546143343154\\
3	-0.638541917945655\\
4	-0.888844614873088\\
5	-1.66978917048705\\
6	-1.65627460452394\\
7	-1.74596807290133\\
8	-1.82660373695372\\
9	-1.55951637013496\\
10	-0.920809945681723\\
11	0.609983494822335\\
12	-3.35665411020363\\
13	-13.8334539879417\\
14	-10.6702625635011\\
15	-8.82676298010207\\
16	-10.1460306344354\\
17	-13.0963570188384\\
18	-15.9027006937911\\
19	-16.983185162852\\
20	-15.7813640046411\\
21	-12.9768951851779\\
22	-9.52754345723547\\
23	-5.82547067929471\\
24	-2.90694314259261\\
25	-1.10159764488429\\
26	-0.302702177193196\\
27	1.48005495672111\\
28	5.38816372598295\\
29	9.8359782594142\\
30	13.1604356537964\\
31	15.1424054271094\\
32	15.2505335937826\\
33	13.1674009758328\\
34	9.89602242856019\\
35	6.65776981404087\\
36	4.19290464287364\\
37	2.55767676785558\\
38	1.57658741417167\\
39	0.982958161595017\\
40	0.584880011130084\\
41	0.32746867900128\\
42	0.177351310429457\\
43	0.0957056861244885\\
44	0.048829931714283\\
45	0.017752075208914\\
46	-0.00456609924085837\\
47	-0.0247940224066794\\
48	-0.0432349160223885\\
49	-0.0582849791540367\\
50	-0.0686807760747858\\
51	-0.0745843779591564\\
52	-0.0771001500484508\\
53	-0.077982895131431\\
54	-0.0791420121758848\\
55	-0.0797401285733787\\
56	-0.077668542064203\\
57	-0.072454393823682\\
58	-0.0652571972693552\\
59	-0.0575648372163458\\
60	-0.0503918900458346\\
61	-0.0441741225666794\\
};
\end{axis}
\end{tikzpicture}%
\caption{Evolution of the system state $x_3(k)$}
\label{subfig:state3}
\end{figure}
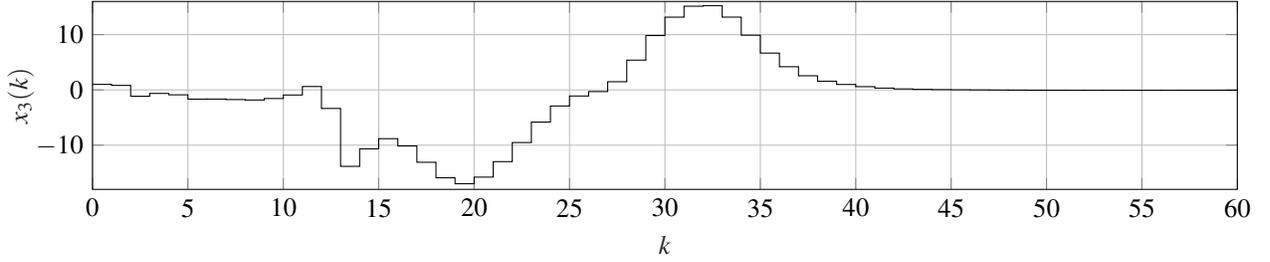

In the following, we show the performance of the proposed algorithm for a single-link robot arm with a DC motor \cite{Carlos2017}. 
After an Euler-forward discretization of the model equations with step-size $h = 0.01$, we obtain the  system
\begin{align}
    \begin{split}
    x_1(k + 1) &= x_1(k) + h x_2(k) \\
    x_2(k + 1) &= x_2(k) + h(36.4 x_3(k) -1.7 x_2(k) - 1309 \mathrm{sin}(x_1(k))  ) \\
    x_3(k + 1) &= x_3(k) + h(-1000 x_3(k) - 3.6 x_2(k) + 100 u(k)) 
        \label{eq:Waste2Ener2}
    \end{split}
\end{align}
where the system state $x = [x_1,\quad x_2,\quad x_3]^\top$ characterizes the position, velocity of the robot and the current of the DC motor. 
We assume that the state can be measured, i.e. $y(k)=x(k)$.
The control input $u$ correspond to the input voltage of the DC motor.   
For the estimation scheme, we choose
$c(e)=\|e\|^2$,
$g(x)=\|x\|^2$ ($D(x,y)=\|x-y\|^2$),
$\bar N=10$,
$\bar N_y=2$,
$\bar N_v=2$,
$\phi_\mathrm{v}(v(k),v(k-1))=[v(k),v(k-1)]^\top$,
$\phi_\mathrm{y}(y(k),y(k-1))=[y(k)^\top,y(k-1)^\top]^\top$.
Further, we initialize the estimator with a controllable signal model.
For the control scheme, we choose
$Q=10 I$,
$R=100 I$,
$Q_{N}=10 I$,
$\Gamma(x)=x^T x$,
$\epsilon(k)=\frac{1}{1+10k}$,
$N=15$.
The state evolution for the initial condition $x_0 = [5,-5,1]^\top$
are depicted in Figure \ref{subfig:state1} to \ref{subfig:state3}.

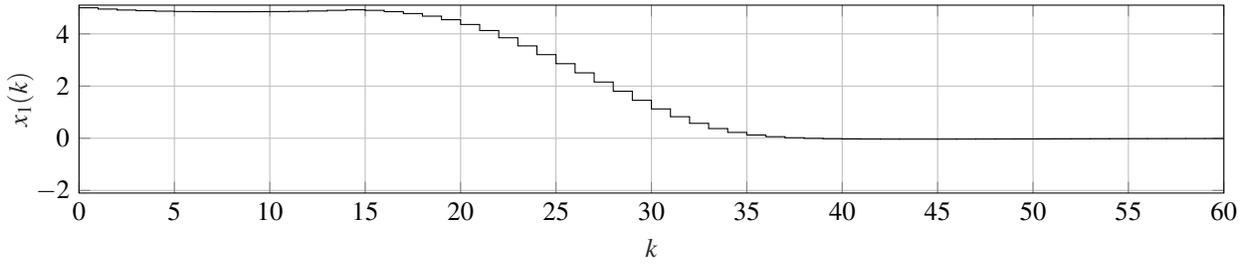
\begin{figure}[t]
\centering
%
%
\begin{tikzpicture}

\begin{axis}[%
width=0.856\columnwidth,
height=2.5cm,
at={(0\columnwidth,0cm)},
scale only axis,
xmin=0,
xmax=60,
xlabel style={font=\color{white!15!black}},
xlabel={$k$},
ymin=-2.1,
ymax=5.1,
ylabel style={font=\color{white!15!black}},
ylabel={$x_1(k)$},
axis background/.style={fill=white},
xmajorgrids,
ymajorgrids,
ylabel style={yshift=-6pt}
]
\addplot[const plot, color=black, forget plot] table[row sep=crcr] {%
0	5\\
1	4.95\\
2	4.91322294795148\\
3	4.88888379286477\\
4	4.86971056909093\\
5	4.85749832364886\\
6	4.85124482755914\\
7	4.84802343253317\\
8	4.84783862361781\\
9	4.8503154680311\\
10	4.85511470782725\\
11	4.86316413414285\\
12	4.87672371937073\\
13	4.90125350836528\\
14	4.92211934106065\\
15	4.90124827356261\\
16	4.85083072919013\\
17	4.77812147737767\\
18	4.67878531713654\\
19	4.54262395987747\\
20	4.3600871088693\\
21	4.12792626593448\\
22	3.85094657616329\\
23	3.53917448409261\\
24	3.20409725140074\\
25	2.8571806489043\\
26	2.50629123907706\\
27	2.15494426895406\\
28	1.80320888417923\\
29	1.45538584258051\\
30	1.12435543364387\\
31	0.825815696041158\\
32	0.572121347120301\\
33	0.371217223035208\\
34	0.224339692817046\\
35	0.124598891817015\\
36	0.0605549156619503\\
37	0.0207044237183557\\
38	-0.00375663110329694\\
39	-0.0186800741117711\\
40	-0.0275769242932598\\
41	-0.032574928748123\\
42	-0.0351085548567896\\
43	-0.0361112633949395\\
44	-0.0361325260633767\\
45	-0.035477185602207\\
46	-0.0343272667218817\\
47	-0.0328103303038978\\
48	-0.0310243602192037\\
49	-0.0290613714068838\\
50	-0.0270077263894761\\
51	-0.0249375657361588\\
52	-0.0229076415774497\\
53	-0.02095753546477\\
54	-0.019113456850084\\
55	-0.0173944967355879\\
56	-0.0158194562893289\\
57	-0.0144036331992964\\
58	-0.0131510435101103\\
59	-0.0120527590377562\\
60	-0.011091312150713\\
61	-0.0102463357555795\\
};
\end{axis}
\end{tikzpicture}%
\caption{Evolution of the system state $x_1(k)$}
\label{subfig:state1}
\end{figure} 
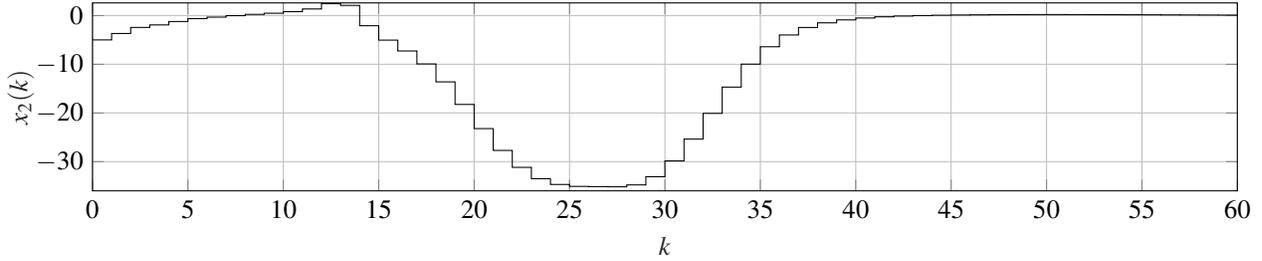
\begin{figure}[t]
\centering
%
%
\begin{tikzpicture}

\begin{axis}[%
width=0.856\columnwidth,
height=2.5cm,
at={(0\columnwidth,0cm)},
scale only axis,
xmin=0,
xmax=60,
xlabel style={font=\color{white!15!black}},
xlabel={ $k$},
ymin=-36,
ymax=2.6,
ylabel style={font=\color{white!15!black}},
ylabel={$x_2(k)$},
axis background/.style={fill=white},
xmajorgrids,
ymajorgrids,
ylabel style={yshift=-6pt}
]
\addplot[const plot, color=black, forget plot] table[row sep=crcr] {%
0	-5\\
1	-3.67770520485169\\
2	-2.43391550867124\\
3	-1.91732237738382\\
4	-1.22122454420724\\
5	-0.62534960897185\\
6	-0.322139502597309\\
7	-0.0184808915354286\\
8	0.247684441328573\\
9	0.479923979614548\\
10	0.804942631560109\\
11	1.35595852278856\\
12	2.45297889945467\\
13	2.08658326953727\\
14	-2.08710674980413\\
15	-5.04175443724803\\
16	-7.27092518124621\\
17	-9.93361602411317\\
18	-13.6161357259066\\
19	-18.253685100817\\
20	-23.2160842934817\\
21	-27.6979689771191\\
22	-31.1772092070677\\
23	-33.5077232691873\\
24	-34.6916602496439\\
25	-35.0889409827242\\
26	-35.1346970122997\\
27	-35.1735384774834\\
28	-34.7823041598716\\
29	-33.1030408936648\\
30	-29.8539737602707\\
31	-25.3694348920857\\
32	-20.0904124085093\\
33	-14.6877530218163\\
34	-9.97408010000304\\
35	-6.40439761550648\\
36	-3.98504919435946\\
37	-2.44610548216527\\
38	-1.49234430084742\\
39	-0.889685018148868\\
40	-0.49980044548632\\
41	-0.253362610866661\\
42	-0.100270853814987\\
43	-0.00212626684371862\\
44	0.0655340461169626\\
45	0.114991888032537\\
46	0.151693641798391\\
47	0.178597008469403\\
48	0.196298881231996\\
49	0.205364501740768\\
50	0.207016065331734\\
51	0.202992415870905\\
52	0.195010611267967\\
53	0.184407861468601\\
54	0.171896011449618\\
55	0.1575040446259\\
56	0.141582309003246\\
57	0.125258968918612\\
58	0.109828447235413\\
59	0.0961446887043177\\
60	0.0844976395133491\\
61	0.0747873975702389\\
};
\end{axis}
\end{tikzpicture}%
\caption{Evolution of the system state $x_2(k)$}
\label{subfig:state2}
\end{figure} 

Overall, our simulation experiments on various examples
show that the proposed  approach performs well in many cases. However,  as also known from adaptive control \cite{And-07},
the state trajectories of unstable systems often show quite a strong peaking behavior.
This is to some extend a fundamental limitation when controlling
unknown unstable systems but more research on this issue is necessary.


\section{Conclusion and Outlook}

Motivation of this research was to develop 
a basic online optimization-based approach
that guarantees convergence for a prototypical problem from adaptive control
and that may serve as a basis for other
online optimization-based (model-free) learning schemes.
To this end, a receding horizon learning scheme
consisting of a receding horizon control scheme and a proximity-based
estimation scheme was proposed. 
For unknown linear time-invariant systems, zero state convergence was proven under rather minimal assumptions.
The motivation to consider linear time-invariant systems stems not only from
the fact that they define an important and tractable benchmark class but 
that they also serve as a measure for the \emph{local} stabilization   of 
unknown nonlinear systems. 
Since the proposed approach relies on predictor maps, it can be considered as an indirect adaptive optimal control method and thus stands in contrast to
direct adaptive optimal control methods such as
reinforcement learning.
From a conceptual point of view, the main ideas
and results of this work were a time-varying model-independent terminal state weighting in the receding horizon control scheme
which does not rely on a controllability assumption (Section \ref{rhcpart}), a convergent proximal estimation scheme that  estimates controllable signal models for predicting the closed loop trajectory (Section \ref{estimationpart})
as well as
a proper combination of the control and estimation scheme to 
achieve guaranteed zero state convergence for completely unknown linear system (Section \ref{overallpart}).


\textcolor{black}{There are several interesting points of future research.
One important question is to investigate under which conditions Assumption \ref{new} b) holds true.
Related to this question is the issue whether or not 
the limiting estimates correctly identify the excited controllable modes of the unknown system
and how fast $\epsilon(k)$ converges to zero
relative to the uncontrollable modes of the system.}
Another interesting issue are constraints. Indeed, constraints satisfaction is impossible without
additional assumptions, but the satisfaction of polytopic input and output constraints, as time goes to infinity  (under some constraint tightening
 or the use (relaxed-)barrier functions
 similarly to \cite{Fel-Ebe-15}),
 should be feasible. 
In particular, such constraints would lead to piecewise quadratic respectively strongly convex value functions of the underlying optimization problems (see Lemma \ref{mp}) and hence it seems reasonable that similar proof arguments
as in this work are applicable.
As mentioned above, shaping the transient behavior is a key  challenge and a well known problem from adaptive control.
The proposed optimization-based formulation 
allows in principle to specify objectives,
constraints (e.g. saturation functions to reduce peaking) and possibly a prior information about the system model in order to improve the transient behavior \cite{And-07}.
Other important research directions are i) to address robustness and uncertainties, for example by using tube techniques \cite{Mayne-05} or techniques from robust control \cite{Sch-18}
or ii) to fit the input-output data not to (lifted) linear signal models but to  nonlinear models such as neural networks
or iii) to study the approach (i.e. the regression) in a Bayesian context.
Moreover, it would be interesting to replace 
in the receding horizon control scheme
the time-varying terminal state weighting with an adaptive one,
in the sense that, for instance, the weight increases only if there is no decrease in the value function after two consecutive time steps.
Finally, despite the involved online optimization is computationally not very demanding, it is important to take real-time aspects into account and to develop anytime iteration schemes or to exploit dual (kernel) formulations 
of the underlying regression problems.


\bibliographystyle{ieeetr}

  \appendix

\section{Proofs}
\label{proofs}

\subsection{Proof of Lemma \ref{mp} and Remark to Assumption \ref{new}}

\textcolor{black}{
Case a) Notice that \eqref{mpc3} can be written
as minimize $~w^\top H w$ subject to $C(k)w=B(k)q(k)$ where $q(k)^\top=[x(k)^\top~r(k)^\top]$,
$w^\top=[\nu_0^\top,\xi_0^\top, ..., \nu_{N-1}^\top,\xi_N^\top]$. 
Since $\xi_N$ is given by $r(k)$, it is easy to see from  \eqref{mpc3} and Assumption \ref{ass1a}
that the solution of \eqref{mpc3} is unique.
Further, since the solution is zero for $x(k)=0,r(k)=0$,
we can write the solution as
 $\xi_i(x(k),p_3(k))=K_{1,i}(k)x(k)+K_{2,i}(k)r(k)$ and 
 $\nu_i(x(k),p_3(k))=K_{3,i}(k)x(k)+K_{4,i}(k)r(k)$
 and the value as $V_3(x(k),p_3(k))=x(k)^\top S_3(k) x(k) + x(k)^\top S_4(k) r(k) + r(k)^\top S_5(k) r(k) \ge 0$ for all $k$ because $Q>0,R>0$.
}

\textcolor{black}{Remark to Assumption \ref{new}. 
Notice that if $\mathrm{rank}[B_0(k),...,B_{N-1}(k)]=n$
for all $k \in \mathbb{N}$ and also
$\mathrm{rank}[\hat B_0,...,\hat B_{N-1}]=n$,
then $C(k)$ and $\lim_{k\rightarrow \infty } C(k)$ have
full column rank, then problem \eqref{mpc3} is feasible for
any $x(k),r(k) \in \mathbb{R}^n$.
Hence, $\tilde w(k) = (C(k)^\top C(k))^{-1}C(k)^\top B(k) q(k)=\tilde K(k)q(k)$ is a suboptimal (feasible) solution of
problem \eqref{mpc3} (i.e. minimize $~w^\top H w$ subject to $C(k)w=B(k)q(k)$). 
Since $\tilde w(k)$ is suboptimal,
$W_3(x(k),p_3(k),\tilde w(k)):=x(k)^\top \tilde S_3(k) x(k) + x(k)^\top \tilde S_4(k) r(k) + r(k)^\top \tilde S_5(k) r(k)$
is an upper bound of the value function.
Further, the limits of
$\{  B(k)\}_{k \in \mathbb{N}}$  and
$\{  (C(k)^\top C(k))^{-1}\}_{k \in \mathbb{N}}$ exist, hence also the limit of
$\lim_{k \rightarrow \infty} \tilde K(k)$
exists. Thus the sequence $\{\tilde K(k)\}_{k \in \mathbb{N}}$ is bounded.
Hence there exists a $M>0$ such that  $\|\tilde S_i(k)\| \le M$, $i=3,4,5$, $\|\tilde K_{j,i}(k)\| \le M$, $i=0...N,j=1...4$, $k \in \mathbb{N}$.
Since $W_3(x(k),p_3(k),\tilde w(k)) \ge V_3(x(k),p_3(k))$ for all $k,x(k),r(k)$,
it follows  that  $\| S_i(k)\|$, $i=3,4,5$, $\| K_{j,i}(k)\|$, $i=0...N,j=1...4$ are uniformly bounded and thus Assumption 5 a) holds true.
}

Case b) Notice that the solution of \eqref{mpc2} is not unique.
Thus we consider the least norm solution of \eqref{mpc2}, i.e.
let $\{\tilde \nu_i(x(k),p_2(k))\}_{i=0}^{N-1}$, $\{\tilde \xi_i(x(k),p_2(k))\}_{i=0}^{N-1}$
be any solution of \eqref{mpc2}.
Then the unique solution 
$\{\nu_i(x(k),p_3(k))\}_{i=0}^{N-1}$, $\{\xi_i(x(k),p_3(k))\}_{i=0}^{N-1}$
of \eqref{mpc3} with the constraint
 $r(k)=\xi_N=\tilde \xi_N(x(k),p_2(k))$ 
and $R=Q=I$ is a solution (i.e. the least norm solution) of \eqref{mpc2}. Hence 
the arguments in Case a) can be applied.

\hfill $\Box$


\subsection{Proof of Lemma \ref{convergenceMPC3}}

Since $Q>0$, there exists a $\alpha>0$ such that $V_3(x,p_3) \ge \alpha x^\top x$ for any $p_3$, i.e. $V_3$ is positive definite and radially unbounded.
For the sake of convenience, we use the notation $x:=x(k),x^+:=x(k+1),p_3:=p_3(k),p_3^+:=p_3(k+1),
r=r(k),r^+:=r(k+1)$. 
In the following, we consider the Lyapunov increment 
 \begin{align}
 \begin{split}
 \label{V0}
V_3(x^+,p_3^+)-V_3(x,p_3) &= V_3(x^+,p_3^+)-V_3(x^+,\tilde p_3^+)\\
&+V_3(x^+,\tilde p_3^+)-V_3(x,p_3),
\end{split}
\end{align}
where $\tilde p_3^+=[k+1,~\tilde r^+]$ is defined below.

Step 1. We first consider the term $V_3(x^+,\tilde p_3^+)-V_3(x,p_3)$. 
Let $\{\nu_i(x,p_3)\}_{i=0}^{N-1}$ be the solution of \eqref{mpc3}
 which we denote in the following by
$\{\nu_i\}_{i=0}^{N-1}$
and let $\{\xi_i(x,p_3)\}_{i=0}^{N}$ be the corresponding predicted states.
Define
\begin{align}
\begin{split}
\label{rt}
\tilde r^+=& 
A_{N}(k+1)x^++\sum_{l=0}^{N-1} B_{N-1-l}(k+1) \nu_{l+1}\\ =&P_N(k+1,x^+,\nu_1,...,\nu_N),
\end{split}
\end{align}
hence $\tilde r^+$ is the predicted terminal state 
at time instant $k+1$ using the state predictor matrices at time $k+1$
and the input sequence $\nu_1,...,\nu_{N-1},\nu_N:=\nu_N(x,p_3):=0$ obtained at time $k$. Hence, $\nu_1,...,\nu_{N-1},0$ is, by construction, feasible for the problem \eqref{mpc3}
at time $k+1$ with $(\tilde p_3^+)^\top=[k+1,~(\tilde r^+)^\top]$. 
Notice that by \eqref{p3} it holds $x^+=Ax+B\nu_0=A_1(k)x+B_0(k)\nu_0-e_0(k)$.
Thus, we have for $i \le N-1$
\begin{align}
\label{statet}
\begin{split}
\tilde \xi_i:=\tilde \xi_i(x^+,\tilde p_3^+) 
= P_i(k+1,P_1(k,x,\nu_0)-e_0(k),\nu_1,...,\nu_i) \\
= P_i(k,P_1(k,x,\nu_0),\nu_1,...,\nu_i)   +P_i(k+1,-e_0(k),0,...,0)\\
+P_i(k+1,P_1(k,x,\nu_0),\nu_1,...,\nu_i) 
-P_i(k,P_1(k,x,\nu_0),\nu_1,...,\nu_i)    \\
= \xi_{i+1}(x,p_3)+w_i(x,p_3)
\end{split}
\end{align}
with
$w_i(x,p_3)=P_i(k+1,-e_0(k),0,...,0)+P_i(k+1,P_1(k,x,\nu_0),\nu_1,...,\nu_i) -P_i(k,P_1(k,x,\nu_0),\nu_1,...,\nu_i)$
and where we used linearity and commutativity of $P_i$ (Assumption \ref{ass0} a) and b)).
Specifically, for $w_i$ we have
\begin{align}
\begin{split}
w_i(x,p_3)=-A_i(k+1)e_0(k)\\
+(A_{i}(k+1)-A_i(k))(A_1(k)x+B_0(k)\nu_0)\\
+\sum_{l=1}^{i} (B_{i-l}(k+1) - B_{i-l}(k)) \nu_{l}.
\end{split}
\end{align}
By Assumption \ref{ass0} we can bound
$w_i(x,p_3)$ for $k \rightarrow \infty$ 
as follows.
Step i):
Let $Q(k)$ be positive definite and uniformly bounded in $k$
by $x^\top Q(k) x \le \lambda x^\top x$ 
for all $x,k \in \mathbb{N}$, then it follows from the
Cauchy-Schwarz's and Young's inequality, that for
  any $\rho>0$  we have
\begin{align}
\label{young}
x^\top Q(k) y \le \lambda ( \frac{\rho}{2} x^\top x+\frac{1}{2\rho}y^\top y).
\end{align}
Step ii) Using the fact that $\nu_i(x,p_3)=K_{3,i}(k)x+K_{4,i}(k)r$ 
with bounded $A_i(k),B_i(k)$ and $K_{3,i}(k),K_{4,i}(k)$ (see Lemma \ref{mp} and \textcolor{black}{Assumption \ref{new} a)})
and  using that 
$\| e_0(k)\|^2 \le \omega_1(k)+\omega_2(k) \|x(k)\|^2 + \omega_3(k)  \| \nu_0\|^2$
we obtain (after some elementary calculations) the bound
\begin{align}
\label{bw}
  \|w_i(x,p_3)\|^2 \le \alpha_{i,1}(k) x^\top x + \alpha_{i,2}(k) r^\top r + \alpha_{i,3}(k)
\end{align}
with $\lim_{k \rightarrow \infty} \alpha_{i,j}(k)=0$
for $j=1,2,3$, $i=0...N-1.$
Consider now
\begin{align}
\label{eq2}
\begin{split}
V_3(x,p_3)=\sum_{i=0}^{N-1} \xi_i(x,p_3)^\top Q \xi_i(x,p_3) + \nu_i(x,p_3)^\top R \nu_i(x,p_3) \\
V_3(x^+,\tilde p_3^+,\nu_1(x,r),...,\nu_{N-1}(x,p_3),0)= \\
\sum_{i=0}^{N-1} \tilde \xi_i^\top Q \tilde \xi_i +  \sum_{i=1}^{N-1} \nu_i(x,p_3)^\top R \nu_i(x,p_3) \\
    =  \sum_{i=1}^{N}  (\xi_i(x,p_3)+w_{i-1}(x,p_3))^\top Q (\xi_i(x,p_3)+w_{i-1}(x,p_3)) \\
    +  \sum_{i=1}^{N-1} \nu_i(x,p_3)^\top R \nu_i(x,p_3).
\end{split}
\end{align}

Due to optimality, we have the inequality
 \begin{align}
 \label{V1}
 \begin{split}
V_3(x^+,\tilde p_3^+)-V_3(x,p_3) \\
\le V_3(x^+,\tilde p_3^+,\nu_1(x,p_3),...,\nu_{N-1}(x,p_3),0)-V_3(x,p_3)\\
=-\xi_0(x,p_3)^\top Q \xi_0(x,p_3) -\nu_0(x,p_3)^\top R \nu_0(x,p_3) \\
+ (\xi_N(x,p_3)+w_{N-1}(x,p_3))^\top Q (\xi_N(x,p_3)+w_{N-1}(x,p_3))\\
+ \sum_{i=1}^{N-1}  w_{i-1}(x,p_3)^\top Q w_{i-1}(x,p_3)  + 2 \xi_i(x,p_3)^\top Q w_{i-1}(x,p_3).
\end{split}
\end{align}
Notice that $\xi_0(x,p_3)=x$ and $\xi_N(x,p_3)=r$.
Similarly as above, i.e. using 
\eqref{bw} and 
\eqref{young} and exploiting the fact that
$\xi_i(x,p_3)$ is linear in $x,r$ and with bounded gains $K_{1,i}(k),
K_{2,i}(k)$ (\textcolor{black}{Assumption \ref{new}, a)}), we can upper bound the expressions in the last two lines of
equation \eqref{V1}
by 
 $(\beta_{1}(k)+\beta_4) x^\top x + (\beta_{2}(k)+\beta_5) r^\top r + \beta_{3}(k)$
with $\lim_{k \rightarrow \infty} \beta_{j}(k)=0$
for $j=1,2,3$ and where $\beta_4>0,\beta_5>0$ can be chosen arbitrarily small.
In more detail, we have for example for the expression $\xi_i Q w_{i-1}$ the bound
$\xi_i Q w_{i-1} \le  \frac{\tilde \beta_{4}}{2} \xi_i Q \xi_i + \frac{1}{2\tilde \beta_{4}} w_{i-1} Q w_{i-1}$
with $\tilde \beta_4$ arbitrarily small. Further $\xi_i(x,p_3)$ is linear in $x,r$
and $w_{i-1}$ obeys \eqref{bw}, which yields a bound of type 
$(\beta_{1}(k)+\beta_4) x^\top x + (\beta_{2}(k)+\beta_5) r^\top r + \beta_{3}(k)$
for this expression.
Applying these arguments to each expression and summing up the obtained bounds
 we obtain
 \begin{align}
 \label{V1a}
 \begin{split}
V_3(x^+,\tilde p_3^+)-V_3(x,p_3) \le
-x^\top Q x -\nu_0(x,p_3)^\top R \nu_0(x,p_3) \\
+(\beta_{1}(k)+\beta_4) x^\top x + (\beta_{2}(k)+\beta_5) r^\top r + \beta_{3}(k),
\end{split}
\end{align}
where $\beta_4,\beta_5$ can be chosen arbitrarily small.

Step 2. We next consider $V_3(x^+,p_3^+)-V_3(x^+,\tilde p_3^+)$. 
Firstly, notice that from \eqref{rt} and \eqref{statet}, we have
$\tilde r^+ = \tilde \xi_N = P_N(k,P_1(k,x,\nu_0),\nu_1,...,\nu_N)+w_N(x,p_3)
=P_1(k,P_N(k,x,\nu_0,...,\nu_N-1),\nu_N)+w_N(x,p_3)=
P_1(k,\xi_N,\nu_N)+w_N(x,p_3)=P_1(k,r,0)+w_N(x,p_3)=A_1(k)r+w_N(x,p_3)
$
hence $\tilde r^+$ depends on $x,p_3$ and converges to zero, i.e.
analogous to \eqref{bw}, we have the bound
  $\|w_N(x,p_3)\|^2 \le \alpha_{N,1}(k) x^\top x + \alpha_{N,2}(k) r^\top r + \alpha_{N,3}(k)$
with $\lim_{k \rightarrow \infty} \alpha_{N,j}(k)=0$ for $j=1,2,3$
and because $r$ converges to zero, we also have
\begin{align}
\label{bw2}
  \|\tilde r^+ \|^2 \le \alpha_{N,1}(k) x^\top x + \alpha_{N,2}(k) r^\top r + \tilde \alpha_{N,3}(k)
\end{align}
with $\lim_{k \rightarrow \infty} \tilde \alpha_{N,3}(k)=0$.
Secondly, by Lemma \ref{mp} \textcolor{black}{and Assumption \ref{new} a)}, it follows
$V_3(x,p_3)=x^\top S_3(k) x + x^\top S_4(k) r + r^\top S_5(k) r$
where $S_i(k)$, $i=3,4,5$ are uniformly bounded  for all $k\in \mathbb{N}$.
Hence, 
\begin{align}
\begin{split}
\label{V2x}
V_3(x^+,p_3^+)-V_3(x^+,\tilde p_3^+) = (x^+)^\top S_4(k+1) (r^+-\tilde r^+) \\
+ (r^+)^\top S_5(k+1) r^+ -
(\tilde r^+)^\top S_5(k+1) \tilde r^+.
\end{split}
\end{align}
Since $x^+=Ax+ B \nu_0$
and using again Young's inequality \eqref{young}
with an arbitrary $\bar \rho>0$, we get
 \begin{align}
 \label{V3}
 \begin{split}
V_3(x^+,p_3^+)-V_3(x^+,\tilde p_3^+)  \\
\le 
 \frac{\bar \rho}{2} (Ax+B \nu_0)^\top   (Ax+B \nu_0) + \gamma(x,p_3)
\end{split} 
\end{align}
with  $\gamma(x,p_3)= 
(r^+)^\top S_5(k+1) r^+ - (\tilde r^+)^\top S_5(k+1) \tilde r^+ + \frac{1}{2 \bar \rho}  (r^+-\tilde r^+)^\top S_4(k+1)^\top S_4(k+1)  (r^+-\tilde r^+)$
that is converging to zero for $k$ to infinity.
In particular, it can be bounded by
 $\gamma(x,p_3) \le  \gamma_1(k) x^\top x +  \gamma_2(k) r^\top r$ 
with $\lim_{k \rightarrow \infty} \gamma_i(k)=0$, $i=1,2$,
which follows from 
$\tilde r^+ =r+w_N(x,p_3)$, Lemma \ref{mp}, \textcolor{black}{Assumption \ref{new} a)} and once more by application of \eqref{young}.

Finally (again \eqref{young}), there exists $c>0$ such that 
$ (Ax+B \nu_0)^\top  (Ax+B \nu_0)
 \le 2 c(x^\top x+\nu_0^\top \nu_0)$,
 hence for any $\bar \rho>0$ we have
  \begin{align}
 \label{V3a}
 \begin{split}
V_3(x^+,p_3^+)-V_3(x^+,\tilde p_3^+) \\
\le 
 \bar \rho c (x^\top x+\nu_0^\top \nu_0)  +  \gamma_1(k) x^\top x +  \gamma_2(k) r^\top r.
\end{split} 
\end{align}

Step 3. Plugging in \eqref{V3a} and \eqref{V1a} in \eqref{V1} 
with $\bar \rho$ such that  
$2 \bar \rho < \mu:=\min\{\lambda_{min}(Q),\lambda_{min}(R)\}$
and $\beta_4$ such that $\mu > 2 \beta_4$ we get
  \begin{align}
 \label{lyapinqe}
\begin{split}
V_3(x^+,p_3^+)-V_3(x,p_3) \\
\le - \frac{\mu}{2} (x^\top x+\nu_0(x,p_3)^\top \nu_0(x,p_3)) \\
  +(\beta_{1}(k)+\beta_4) x^\top x + (\beta_{2}(k)+\beta_5) r^\top r \\
  + \beta_{3}(k)
 + \gamma_1(k) x^\top x +  \gamma_2(k) r^\top r.
 \end{split}
\end{align} 
 Since $r=r(k),\gamma_i(k),\beta_i(k)$ converge to zero
 (and thus are bounded sequences)
 and since $x(k)$ is defined for all $k \in \mathbb{N}$, 
 it follows that 
 $x(k)$ converges to zero, i.e
 there exists a time $\tilde k$, such that
 for all $k \ge \tilde k$, the right hand side of 
 \eqref{lyapinqe} is negative for $x(k) \not = 0$. 
 Finally, due to Lemma \ref{mp}, $V_3(0,p_3)=0$, hence
 if $x(k)$ converges to zero then also $u(k)$.  \hfill $\Box$

\subsection{Proof of Lemma \ref{lem1a}}

Proof by contradiction.
Suppose for all $\rho>0$  
there exists an $x \in \mathbb{R}^N$,  $ k \in \mathbb{N}$
and $\epsilon>0$
such that
$x_N(x,p_1)^\top Q_N x_N(x,p_1) > V_2(x,p_2)+\epsilon \rho$.
Then by \eqref{mpc1} and Assumption \ref{ass0}(a),
we have $V_1(x,p_1) > \frac{\Gamma(x) V_2(x,p_2)}{\epsilon}+\rho \Gamma(x)$.
On the other hand we have $V_1(x,p_1,\nu_0(x,p_2),...,\nu_{N-1}(x,p_2))=
 \sum\limits_{i = 0}^{N - 1}  \xi_i(x,p_2)^{\top} Q \xi_i(x,p_2) + \nu_i(x,p_2)^{\top} R \nu_i(x,p_2)  +
\frac{\Gamma(x) V_2(x,p_2)}{\epsilon}$,
where $\{\nu_i(x,p_2)\}_{i=0}^{N-1}$, $\{\xi_i(x,p_2)\}_{i=0}^{N}$ is a solution of \eqref{mpc2} and it is feasible for \eqref{mpc1}.
By optimality, we must have 
$V_1(x,p_1) \le V_1(x,p_1,\nu_0(x,p_2),...,\nu_{N-1}(x,p_2))$, hence

\begin{align}
\begin{split}
\label{contra1}
  \frac{\Gamma(x) V_2(x,p_2)}{\epsilon}+\rho \Gamma(x) 
  < \frac{\Gamma(x) V_2(x,p_2)}{\epsilon}\\
  + \sum\limits_{i = 0}^{N - 1}  \xi_i(x,p_2)^{\top} Q \xi_i(x,p_2) + \nu_i(x,p_2)^{\top} R \nu_i(x,p_2) 
  \end{split}
  \end{align}
  must hold.
However, \eqref{contra1} cannot be true for all $\rho>0$
since  $\Gamma(x) \ge c (\sum_{i=0}^{N-1} \|\xi_i(x,p_2)\|^2+\|\nu_i(x,p_2)\|^2)$ and
therefore there exists a $\rho >0$ (independent of $x$, $p_2$) such that $\rho \Gamma(x) \ge 
\rho c (\sum_{i=0}^{N-1} \|\xi_i(x,p_2)\|^2+\|\nu_i(x,p_2)\|^2)
\ge
\sum\limits_{i = 0}^{N - 1}  \xi_i(x,p_2)^{\top} Q \xi_i(x,p_2) + \nu_i(x,p_2)^{\top} R \nu_i(x,p_2)$.
\hfill $\Box$

\subsection{Proof of Theorem \ref{MPC1vsMPC3}}

To prove the result, we use Lemma \ref{convergenceMPC3}.
We first show the following: 
\emph{ There
exists an (appropriately constructed) sequence $\{p_3(k):=\tilde p_1(k)^\top=[k,r(k)^\top]\}_{k \in \mathbb{N}}$, such that
 the solution of \eqref{mpc1}  at time instant $k$ with $p_1=p_1(k)$
 and $x=x(k)$, where $x(k)$ is the state of the closed loop \eqref{mpc1},\eqref{sysmpc},
 is equivalent to the solution of \eqref{mpc3} for $x=x(k)$,
 $p_3=p_3(k)$ and thus the state sequence of the closed loop \eqref{mpc1},\eqref{sysmpc}
 coincides with that of \eqref{mpc3},\eqref{sysmpc}.
 Moreover we show that the constructed sequence
 $\{r(k)\}_{k \in \mathbb{N}}$ converges to zero.}

Step 1. Let $\{u_i(x,p_1)\}_{i=0}^{N-1}$, $\{x_i(x,p_1)\}_{i=0}^{N}$ be the unique solution of \eqref{mpc1}.
Then due to \eqref{predictor}, it is also the unique solution of 
\begin{align}
\label{mpc1a}
\begin{split}
     \tilde V_1(x,\tilde p_1)=\min  ~&~  \sum\limits_{i = 0}^{N - 1}  \xi_i^{\top} Q \xi_i + \nu_i^{\top} R \nu_i \\
    \text{s.t. } 
    ~&~ \xi_{i + 1} = A_{i+1}(k)x+ \sum_{l=0}^iB_{i-l}(k) \nu_l,  \\
                     ~&~ \xi_0 = x, ~i=0...N-1, \\
                 ~&~ \xi_N = x_N(x,p_1)
\end{split}    
\end{align}
with $\tilde p_1^\top=[k,x_N(x,p_1)^\top]$.
Thus, any closed loop trajectory of \eqref{sysmpc} and \eqref{mpc1}
coincides with a closed loop trajectory of \eqref{sysmpc} and \eqref{mpc1a}
with the same initial data.
Since \eqref{mpc1a} is an instance of \eqref{mpc3} with $r=x_N(x,p_1)$
\textcolor{black}{and thus Assumption \ref{new} b) implies Assumption \ref{new} a)}, it remains
to show (in Step 2a and 2b below) that $r(k):=x_N(x(k),p_1(k))$ converges to zero as $k$ goes to infinity in order to apply Lemma \ref{convergenceMPC3}.

Step 2. We next apply Lemma \ref{lem1a}. Therefore,
we first show that there exists a $\tilde c>0$ 
such that for any $k \in \mathbb{N}$ ($p_2(k)=k$)
and any $x(k)$ we have 
$\tilde c \alpha x(k)^\top x(k) \ge \sum_{i=0}^{N-1} \|\xi_i(x(k),p_2(k))\|^2+\|\nu_i(x(k),p_2(k))\|^2$,
where $\{\nu_i(x(k),p_2(k))\}_{i=0}^{N-1}$, $\{\xi_i(x(k),p_2(k))\}_{i=0}^{N-1}$
is some solution of \eqref{mpc2} at a time step $k$. 

Notice that $\eqref{mpc2}$ has not necessarily a unique solution
and Lemma \ref{lem1a} refers to \emph{some} solution.
Therefore, according to Lemma \ref{mp}, there is a solution
$\{ \xi_i(x(k),p_2(k))\}_{i=0}^{N-1}$, $\{ \nu_i(x(k),p_2(k))\}_{i=0}^{N-1}$
of \eqref{mpc2} (i.e. the least norm solution) which is linear in $x$
and uniformly bounded in $k$, i.e. $\xi_i(x(k),p_2(k))=K_{1,i}(k)x(k)$,
$\nu_i(x(k),p_2(k))=K_{2,i}(k)x(k)$.

Consequently, there exist $c_1>0$ such that for $i=0,...,N$
$\| \xi_i(x(k),p_2(k)) \|^2 \le c_1 x(k)^\top x(k)$,
$\| \nu_i(x(k),p_2(k)) \|^2 \le c_1 x(k)^\top x(k)$.
Hence, we have
$\sum\limits_{i = 0}^{N - 1}  \| \xi_i(x(k),p_2(k)) \|^2 + \| \nu_i(x(k),p_2(k)) \|^2 
\le \tilde c \alpha x(k)^\top x(k)$ for some $\tilde c >0$ 
and we can finally apply Lemma \ref{lem1a} to get $x_N(x(k),p_1(k))^\top Q_N x_N(x(k),p_1(k))   \le V_2(x(k),p_2(k))+\epsilon(k)\rho$. 

Step 3. Since $Q_N>0$ and $\{\epsilon(k)\}_{k \in \mathbb{N}}$ converges to zero, it remains to show  that $V_2(x(k),p_2(k))$ defined in \eqref{mpc2} converges to zero,
which however is a direct consequence of Assumption \ref{ass0}(c). 
Hence, $r(k):=x_N(x(k),p_1(k))$ converges to zero.

Summarizing, we have shown that there
exists a sequence $\{p_3(k):=\tilde p_1(k)=[k,x_N(x(k),p_1(k))^\top]^\top\}_{k \in \mathbb{N}}$ such that
 the solution of the closed loop  \eqref{sysmpc},\eqref{mpc1} is equivalent to the solution of  the closed loop \eqref{sysmpc}, \eqref{mpc3}
 and  $\{\tilde p_1(k)\}_{k \in \mathbb{N}}$ converges to zero.
 Thus, by Lemma \ref{convergenceMPC3}, the state and input sequence of the closed loop system \eqref{sysmpc} and \eqref{mpc1}
 converges to zero. 

 \hfill $\Box$

\subsection{Proof of Lemma \ref{timevaryingmin}}

From the first order optimality condition for the right hand side of \eqref{eq:prox_iter2}, we get
\begin{align}
 \nabla_x f(x_{k+1}^*,k) =- \nabla_{\mathrm{x}} D(x_{k+1}^*,\tilde x_k). \label{eq:nec_opt}
\end{align}
Moreover, due to the convexity of $f$, it holds for all $x \in \mathbb{R}^n$
\begin{align}
f(x,k) \geq f(x^*_{k + 1},k) + \nabla_x f(x^*_{k + 1},k)^{\top} (x - x^*_{k + 1}). \label{eq:taylor}
\end{align}
Evaluating \eqref{eq:taylor} at a time invariant minimizer $x^* \in X$ and inserting \eqref{eq:nec_opt} into \eqref{eq:taylor}, we  get
\begin{align}
f(x^*,k) - f(x^*_{k + 1},k) \geq - \nabla_{\mathrm{x}} D(x^*_{k + 1},\tilde x_k) (x^* - x^*_{k + 1}). \label{eq:taylor_D}
\end{align}
Inserting the definition of the Bregman distance and the gradient of the Bregman distance with $\nabla_x D(x,y) = \nabla g(x) - \nabla g(y)$ into \eqref{eq:taylor_D} yields 
\begin{align}
\label{threepoint}
\begin{split}
\hspace{-2.645mm}f(x^*,k) - f(x^*_{k + 1},k)  \geq
[\nabla g(x^*_k) - \nabla g(x^*_{k + 1})]^{\top} (x^* - x^*_{k + 1})\\
= D(x^*,x^*_{k + 1}) + D(x^*_{k + 1},\tilde x_k)- D(x^*,\tilde x_k),
\end{split}
\end{align}
where the last equality follows from the three-point identity of 
the Bregman distance (see also \cite[Proof of Proposition 3.6]{Cen-92}). 
Since $D$ is convex in the second argument, we have
\begin{align}
\label{l1}
\begin{split}
     D(x^*,\tilde x_{k+1}) &\le (1-\lambda_{k+1})D(x^*,x^*_{k+1})+\lambda_{k+1} D(x^*,\tilde x_k) \\
           & \le  (1-\lambda_{k+1})[D(x^*,\tilde x_k)-D(x^*_{k+1},\tilde x_k)\\
           &~~+f(x^*,k)-f(x^*_{k+1},k)] + \lambda_{k+1} D(x^*,\tilde x_k), 
\end{split}
\end{align}
and hence
\begin{align}
\label{l2}
\begin{split}
     D(x^*,\tilde x_{k+1})         - D(x^*,\tilde x_k) \\
     \le  (1-\lambda_{max})[-D(x^*_{k+1},\tilde x_k)
           +f(x^*,k)-f(x^*_{k+1},k)].
\end{split}
\end{align}
Based on \eqref{l2}, we now utilize $V(x)=D(x^*,x)$ as a Lyapunov-like function.
In particular, $V$ is nonnegative, zero if and only if $x=x^*$
and strictly convex and hence radially unbounded. 
\textcolor{black}{Since the right hand side of  \eqref{l2} is non-positive,  $\tilde x_k$ is a bounded sequence.}
Moreover, $V$ is strictly monotonically decreasing as long
as i) $D(x^*_{k+1},\tilde x_k)>0$ and ii) $f(x^*,k)-f(x^*_{k+1},k)<0$.
Hence, we infer that 
$D(x^*_{k+1},\tilde x_k)$ must converge to zero, and we have
$\lim_{k \rightarrow  \infty} \|x^*_{k+1}-\tilde x_k\|=0$.
By $\tilde x_{k+1} = (1-\lambda_{k+1}) x^*_{k+1}+\lambda_{k+1} \tilde x_k$,
we  also have $\lim_{k \rightarrow  \infty} \|\tilde x_{k+1}-\tilde x_k\|=0$.
Consequently $\tilde x_k$ converges to a point,
say $\tilde x^0$.
In addition, due to 
$\lim_{k \rightarrow  \infty} \|\tilde x_{k+1}-\tilde x_k\|=0$
and $\lambda_k \le \lambda_{max}<1$, it must also hold that 
 $x^*_k$ converges to $\tilde x^0.$
Finally, also $f(x^*,k)-f(x^*_{k+1},k)$ must converge to zero,
which implies that $f(\tilde x^0,k)=f(x^*,k)=0$.

\hfill $\Box$

\subsection{Proof of Lemma \ref{staycontrollable}}

Let $C(\lambda)=[B(\lambda), A(\lambda)B(\lambda),...,A^{n-1}(\lambda)B(\lambda)]$
be the controllability matrix and let $p(\lambda)=\mathrm{det(C(\lambda) C(\lambda)^\top})$. Then $p(\lambda)$ is a polynomial of at most degree $2n^2$
with $p(0)=0$ and $p(1)>0$. Hence, $p$ is not identical zero and 
there exist at most $2n^2$ real zeros of $p$ over any interval. 
Therefore, if $p$ is evaluated at $2n^2+1$ distinct points, then there is at least
one point where $p$ is nonzero. The claim follows from this basic observation.
\hfill $\Box$ 

\subsection{Proof of Theorem \ref{xxx}}

(i) By Assumption \ref{zerovalue} and \ref{exo}   and by eliminating the variable $e$ in \eqref{sp1}, we can apply Lemma \ref{timevaryingmin}. Hence,
we have 
$\lim_{k \rightarrow \infty}\theta(k)=\theta^*$.

(ii) From the construction \eqref{predictormaps} of the predictor maps
and the fact that $s(k)-R(k)\theta^*(k)=e^*(k)$ corresponds to the linear
system of equations 
\begin{align}
\label{help}
x(j)=A(k)x(j-1)+B(k)u(j-1)+e^*_{j-k+\bar N}(k),
\end{align}
 $j=k...k-\bar N+1$, 
$e^*(k)=[e^*_{\bar N}(k)^\top \ldots e^*_{1}(k)^\top]$,
as well as from the convergence property (i), i.e. by $\lim_{k \rightarrow \infty }A(k) = \hat A$,
$\lim_{k \rightarrow \infty }B(k) = \hat B$,
 property \eqref{p2}, \eqref{p4} in Assumption \ref{ass0} is satisfied.
It remains to show property \eqref{p3}.
Notice that by \eqref{help}, we have for any $k \in \mathbb{N}$
and $j=0...\bar N-1$
\begin{align}
\begin{split}
x(j+k-\bar N+1)=A_{j}(k) x(k-\bar N+1) \\
+\sum_{l=0}^{j-1} B_{j-1-l}(k) u(l+k-\bar N+1)
+\sum_{l=0}^{j-1} A(k)^{j-1-l} e_{l+1}^*(k)
\end{split}
\end{align}
Since 
$\lim_{k \rightarrow \infty }A_i(k) = \hat A_i$,
$\lim_{k \rightarrow \infty }B_i(k) = \hat B_i$,
we have
\begin{align}
\begin{split}
x(j+k-\bar N+1)=\hat A_{j} x(k-\bar N+1) \\
+\sum_{l=0}^{j-1} \hat B_{j-1-l} u(l+k-\bar N+1)
+\sum_{l=0}^{j-1} A(k)^{j-1-l} e_{l+1}^*(k) \\
 + 
(A_{j}(k)-\hat A_j) x(k-\bar N+1) \\
+\sum_{l=0}^{j-1} (B_{j-1-l}(k)-\hat B_{j-l-1}) u(l+k-\bar N+1).
\end{split}
\end{align}

By Lemma \ref{timevaryingmin}, $\lim_{k \rightarrow \infty }e^*_{l+1}(k)=0 $, $l=0...\bar N-1$, hence there exists a sequence
$\{\omega_1(k)\}_{k \in \mathbb{N}}$ that converges to zero
and such that for all $j=0...\bar N-1$ it holds:
$\|\sum_{l=0}^{j-1} A(k)^{j-1-l} e_{l+1}^*(k)\|^2 \le \omega_1(k)$.

By Cauchy-Schwarz's and Young's inequality, there exist sequences
$\{\omega_2(k)\}_{k \in \mathbb{N}}$, $\{\omega_3(k)\}_{k \in \mathbb{N}}$ that both converge to zero
and such that $ \|(A_{j}(k)-\hat A_j) x(k-\bar N+1) \|^2 \le \omega_2(k) 
\|x(k-\bar N+1)\|^2$
and such that for all $j=0...\bar N-1$
$\|\sum_{l=0}^{j-1} (B_{j-1-l}(k)-\hat B_{j-l-1}) u(l+k-\bar N+1)\|^2
  \le \omega_2(k)  \sum_{l=0}^{j-1} \|u(l+k-\bar N+1)\|^2$.
Hence
\begin{align}
\label{dum}
\begin{split}
x(j+k-\bar N+1)&=\hat A_{j} x(k-\bar N+1) +e_j(k-\bar N+1)\\
&+\sum_{l=0}^{j-1} \hat B_{j-1-l} u(l+k-\bar N+1)
\end{split}
\end{align}
where $e_j(k-\bar N+1)=
\sum_{l=0}^{j-1} A(k)^{j-1-l} e_{l+1}^*(k) 
+ (A_{j}(k)-\hat A_j) x(k-\bar N+1) 
+\sum_{l=0}^{j-1} (B_{j-1-l}(k)-\hat B_{j-l-1}) u(l+k-\bar N+1)$, $j=0...\bar N-1$, satisfies the error bounds in Assumption \ref{ass0}(b) as shown above.
In more detail, since \eqref{dum} holds for any $k \in \mathbb{N}$, let $k=\tilde k + \bar N-1$, then we get the desired asymptotically correct prediction property \eqref{p3} 
with respect to \eqref{eq:exo} with $N=\bar N-1$:
\begin{align}
\begin{split}
\hat A_{j} x(\tilde k) 
+\sum_{l=0}^{j-1} \hat B_{j-1-l} u(l+\tilde k)+e_j(\tilde k)\\
=x(j+\tilde k) = A^{j} x(\tilde k)+\sum_{l=0}^{j-1} A^{j-1-l} B u(l+\tilde k).
\end{split}
\end{align}

(iii) Equation \eqref{stabilizability} follows directly from Lemma \ref{staycontrollable}. In particular, 
suppose $\tilde \theta(k-1)^\top = [\mathrm{vec}(A(k-1))^\top,~\mathrm{vec}(B(k-1))^\top]$ is
controllable (which holds for $k=1$) and suppose $(A_u(k),B_u(k))$ obtained from
$\theta^*(k)^\top = [\mathrm{vec}(A_u(k))^\top,~\mathrm{vec}(B_u(k))^\top]$
is not controllable.
Then $\lambda_k$ and thus $\tilde \theta(k)^\top = [\mathrm{vec}(A(k))^\top,~\mathrm{vec}(B(k))^\top]$
such that $(A(k),B(k))$ is controllable can be constructed according to Lemma \ref{staycontrollable}. 
(If $\theta^*(k)^\top = [\mathrm{vec}(A_u(k))^\top,~\mathrm{vec}(B_u(k))^\top]$
is controllable, then simply choose $\lambda_k=0$).
\hfill $\Box$

\subsection{Proof of Lemma \ref{finalstep}}

We utilize the Kalman decomposition. Since we consider only input and output data and since Assumption \ref{ass1ground} holds,
we can assume without loss of generality that $(F,G,H)$ in \eqref{sys} is structured as follows
\begin{align}
\label{kalman}
\begin{split}
  \begin{bmatrix}
 z_1(k+1) \\
 z_2(k+1)\\
 z_3(k+1)
 \end{bmatrix}
 = &
 \begin{bmatrix}
  F_1 & F_2 & 0 \\
  0 & F_3 & 0 \\
  F_4 & F_5 &  F_6 \\
 \end{bmatrix}
 \begin{bmatrix}
 z_1(k) \\
 z_2(k)\\
 z_3(k)
 \end{bmatrix}+
 \begin{bmatrix}
  G_1 \\
  0  \\
  G_2  \\
 \end{bmatrix}
 v(k) \\
 y(k) =& \begin{bmatrix} H_1 & H_2 & 0 \end{bmatrix}
  \begin{bmatrix}
 z_1(k) \\
 z_2(k)\\
 z_3(k)
 \end{bmatrix},
 \end{split}
\end{align}
where $F_3,F_6$ are stable (eigenvalues are in the interior of the complex unit disc) and the subsystem
\begin{align}
\label{subsys}
\begin{split}
  \begin{bmatrix}
 z_1(k+1) \\
 z_2(k+1)
 \end{bmatrix}
 = &
 \begin{bmatrix}
  F_1 & F_2  \\
  0 & F_3  
 \end{bmatrix}
 \begin{bmatrix}
 z_1(k) \\
 z_2(k)
 \end{bmatrix}+
 \begin{bmatrix}
  G_1 \\
  0  
 \end{bmatrix}
 v(k) \\
 :=& F_s
z_s (k) +
 G_s v(k)
 \\
 y(k) =& \begin{bmatrix} H_1 & H_2 \end{bmatrix} z_s(k) := H_s z_s(k)
 \end{split}
\end{align}
is observable and stabilizable. 
Due to the cascaded structure, it follows that if the state and input
of the subsystem \eqref{subsys} goes to zero, then also the state (and the input) of
the overall system \eqref{kalman}, since $F_6$ is stable.
Notice further that the output sequences of
the subsystem  are equivalent to the output sequences
of the overall system (for the same input sequences).
Since the subsystem is observable and $m \ge n$ is known, it follows that if 
$x(k)=\phi_{\mathrm{y}}(y(k),...,y(k-m+1))=
 \begin{bmatrix}
    y(k)^\top,...,y(k-m+1)^\top
 \end{bmatrix}^\top$ and
 $u(k)=\phi_{\mathrm{v}}(v(k),...,v(k-m+1))=[v(k)^\top,...,v(k-m+1)^\top]^\top$ 
goes to zero, then also 
$\{v(k)\}_{k \in \mathbb{N}}$,
$\{y(k)\}_{k \in \mathbb{N}}$ of the subsystem \eqref{subsys} (as well
as of the overall system \eqref{kalman}).
As a final step, consider 
$ x(k)=\phi_{\mathrm{y}}(y(k),...,y(k-m+1))=
 \begin{bmatrix}
    y(k)^\top,...,y(k-m+1)^\top
 \end{bmatrix}^\top$ which is given by
\begin{align}
\label{stack}
\begin{split}
\begin{bmatrix} 
H_s F_s^{m-1} z_s(k-m+1) + H_s \sum_{l=0}^{m-2} F_s^{m-2-l} G_s v(k-m+1+l) \\
 \vdots \\
 H_s F_s z_s(k-m+1)+ H_s G_s v(k-m+1)\\
 H_s z_s(k-m+1)
\end{bmatrix}.  
\end{split}
\end{align}
Due to \eqref{arx}, \eqref{stack} can be compactly written as $x(k)  = O z_s(k-m+1) + R u(k-1)$.
Since \eqref{subsys} is observable, the (observability) matrix $O$ has full column rank,
hence we have
$z_s(k-m+1)=(O^\top O)^{-1} O^\top x(k) -(O^\top O)^{-1} O^\top R u(k-1)$.
Thus 
\begin{align}
\label{hsm}
\begin{split}
x(k+1) & = O z_s(k-m+2) + R u(k) \\
& = O F_s z_s(k-m+1) + O G_s v(k-m+1) + R u(k) \\
& = O F_s ((O^\top O)^{-1} O^\top x(k) -(O^\top O)^{-1} O^\top R u(k-1))\\
&+O G_s v(k-m+1) + R u(k) 
\end{split}
\end{align}
which can be written as $x(k+1)=A x(k)+ Bu(k)$.
Consequently 
$\{u(k)\}_{k \in \mathbb{N}}$,
$\{x(k)\}_{k \in \mathbb{N}}$ satisfy Assumption \ref{exo}.
To show that $(A,B)$ is stabilizable, choose $v(k)=K_s z_s(k)$
such that $F_s+G_sK_s$ is stable, which is possible since
\eqref{subsys} is stabilizable. Then this choice implies that 
\eqref{hsm} is stable, 
which follows directly from \eqref{stack}.
In more detail, from \eqref{hsm}
we have
$w_{k+1}=(F_s+G_s K_s)w_k$ with $w_{k}:=O^Tx_k$. Therefore $w_k$
goes to zero and thus $x_k$, hence
we have shown that the obtained pair $(A,B)$ is stabilizable.

Concerning Remark  \ref{intchain}. Notice that the input vector $u(k)$ 
contain past values of the actual input $v(k)$. These past input values,
however, can be eliminated by augmenting
additional state variables
to \eqref{hsm} and by defining an integrator chain
dynamics of the form $\zeta_1(k+1)=\zeta_2(k),...,
\zeta_{m-1}(k+1)=\zeta_{m}(k), \zeta_{m}(k+1)=v(k)$,
hence we have $\zeta_1(k)=v(k-m)$ etc.
Stabilizability of the augmented system follows from the stabilizability
of $(A,B)$, since the states of the integrator chain converge to zero,
if $v(k)$ converges to zero, i.e. a stabilizing feedback is again $v(k)=K_s z_s(k)$.
\hfill $\Box$

\end{document}